\documentclass[12pt]{article}
\pdfoutput=1

\usepackage{draft} 
\usepackage{hyperref}
\usepackage{graphicx,color,subfig}
\usepackage{cite}
\usepackage{mciteplus}
\usepackage{skak}
\usepackage{empheq}
\usepackage{anyfontsize}
\usepackage{slashbox}
\DeclareFontFamily{OT1}{pzc}{}
\DeclareFontShape{OT1}{pzc}{m}{it}{<-> s * [1.10] pzcmi7t}{}
\DeclareMathAlphabet{\mathpzc}{OT1}{pzc}{m}{it}

\def\be#1\ee{\begin{align}#1\end{align}}

\begin{document}

\unitlength = .8mm

\begin{titlepage}

\begin{center}

\hfill \\
\hfill \\
\vskip 1cm

\title{Strings in Ramond-Ramond Backgrounds from the Neveu-Schwarz-Ramond Formalism}

\author{Minjae Cho$^\spadesuit$, Scott Collier$^\spadesuit$, Xi Yin$^\spadesuit{}^\diamondsuit$}

\address{
$^\spadesuit$Jefferson Physical Laboratory, Harvard University, \\
Cambridge, MA 02138 USA
\\
$^\diamondsuit$Center for Theoretical Physics, Massachusetts Institute of Technology, \\
Cambridge, MA 02139 USA
}

\email{minjaecho@fas.harvard.edu, scollier@g.harvard.edu, xiyin@g.harvard.edu}

\end{center}

\abstract{ We treat RR flux backgrounds of type II string theory in the framework of closed superstring field theory based on the NSR formalism, focusing on two examples: (1) the pp-wave background supported by 5-form flux, and (2) $AdS_3\times S^3\times M_4$ supported by mixed 3-form fluxes. In both cases, we analyze the classical string field solution perturbatively, and compute the correction to the dispersion relation of string states to quadratic order in the RR flux. In the first example, our result is in a delicate way consistent with that obtained from lightcone quantization of the Green-Schwarz string. In the second example, we will obtain numerically the mass corrections to pulsating type IIB strings in $AdS_3\times S^3\times M_4$. Our results, valid at finite $AdS$ radius, agree with previously known answers in the semiclassical limit and in the BMN limit respectively.
}

\vfill

\end{titlepage}

\eject

\begingroup
\hypersetup{linkcolor=black}

\tableofcontents

\endgroup

\section{Introduction} 

The Neveu-Schwarz-Ramond (NSR) formulation \cite{Friedan:1985ge,Friedan1985,Witten:2012bh,Witten:2013cia,Sen:2015hia,polchinski1998string} of the perturbation theory of superstrings is based on a worldsheet $(1,1)$ superconformal field theory coupled to the $b,c,\B,\C$ ghost system, with total central charge zero, and a BRST symmetry such that the superconformal currents are the BRST transformation of $b$ and $\B$ ghosts. This framework can be used to formulate the perturbation theory of type II superstrings propagating in a spacetime background that corresponds to a solution of the supergravity equations of motion in which the NSNS fields are turned on. A deformation of the NSNS background corresponds to deforming the worldsheet SCFT by a marginal primary $V_{\rm NS}^{0,0}$, known as the NSNS vertex operator in $(0,0)$-picture. Generic supergravity solutions also involve nonzero RR field strengths (fluxes), and the latter play key roles in string compactifications and holographic dualities \cite{Kachru2003a,Maldacena1999e,David2002}. As the RR vertex operators come with half-integer picture numbers, it does not seem sensible to deform the worldsheet SCFT by the RR vertex operators. This is often thought of as a fundamental limitation of the NSR formalism, and alternative formalisms were deemed necessary \cite{Berkovits1999,Berkovits1999a,Berkovits2000b,Berkovits2000a,Berkovits2000,Berkovits2001}.

It was suggested in the seminal work of Friedan, Martinec, and Shenker \cite{Friedan:1985ge} that the fermionic contributions to supergravity equations can be recovered by the inclusion of an even number of Ramond-sector vertex operator insertions. This idea was investigated in detail two decades ago by Berenstein and Leigh \cite{Berenstein:1999jq,Berenstein:1999ip}. It was proposed that the NSR formalism can be extended to RR backgrounds by deforming the worldsheet action by
\ie\label{actdef}
\Delta S = \int d^2z \, {\cal V}^{0,0}_{\rm NS}(z,\bar z) + \int d^2z \, {\bf P}_{1\over 2} {\bf \widetilde P}_{1\over 2} {\cal V}_{\rm R}^{-{1\over 2},-{1\over 2}}(z,\bar z),
\fe
where ${\cal V}^{0,0}_{\rm NS}$ stands for an NSNS vertex operator in the $(0,0)$ picture, and ${\cal V}_{\rm R}^{-{1\over 2},-{1\over 2}}$ stands for an RR vertex operator in the $(-{1\over 2}, -{1\over 2})$ picture. The formal object ${\bf P}_{1\over 2}$ was introduced in \cite{Berenstein:1999jq,Berenstein:1999ip} such that a pair of ${\bf P}_{1\over 2}$ insertions is to be thought of as a single picture changing operator (PCO) ${\cal X}$ (and similarly for the anti-holomorphic object ${\bf \widetilde P}_{1\over 2}$).
The obvious difficulty with this prescription is that ${\bf P}_{1\over 2}$ is not well defined, as the locations of PCOs are unspecified. While in a BRST invariant correlator the locations of PCOs are unimportant \cite{Friedan:1985ge}, in superstring perturbation theory the moduli space integrand is typically only BRST invariant up to total derivatives.

One may attempt to make (\ref{actdef}) precise by making a specific choice of the location of PCOs, and replace the insertion of $\exp(-\Delta S)$ with
\ie\label{rrdef}
{}&\exp\left[ - \int d^2z\, {\cal V}^{0,0}_{\rm NS}(z,\bar z) \right] \sum_{n=0}^\infty {1\over (2n)!} \left[ \int d^2z \, {\cal V}_{\rm R}^{-{1\over 2},-{1\over 2}}(z,\bar z) \right]^n \left[ \int d^2z \, {\cal V}_{\rm R}^{{1\over 2},{1\over 2}}(z,\bar z) \right]^n,
\fe
where ${\cal V}_{\rm R}^{{1\over 2},{1\over 2}}$ is related to ${\cal V}_{\rm R}^{-{1\over 2},-{1\over 2}}$ by picture raising. From the worldsheet CFT point of view, such a deformation is nonlocal, but nonetheless Weyl invariance ought to be preserved. Indeed, it was shown in \cite{Berenstein:1999jq,Berenstein:1999ip} that the supergravity equations involving RR fields can be recovered from the cancellation of the Weyl anomaly on the worldsheet.
It is not a priori clear, however, how to define (\ref{rrdef}) precisely beyond leading order in the deformation parameter, as the vertex operators in question are expected to be off-shell.

The closed superstring field theory (SFT) based on the NSR formalism \cite{Zwiebach:1992ie, Pius:2014iaa, Sen:2014pia, Sen:2014dqa, Sen:2015hha, Sen:2016uzq, deLacroix:2017lif} is a framework that in principle provides a systematic formulation of superstring perturbation theory in any closed string background, at the classical as well as the quantum level. While the NSNS and RR deformation operators in (\ref{rrdef}) can loosely be identified with string fields to leading orders, in the SFT approach, the background deformation is entirely captured by a solution to the string field equation which lives in the Hilbert space of the original, undeformed, worldsheet CFT. 

In this paper we will adopt the classical SFT framework to analyze RR background deformations and the string spectrum. Our general strategy is laid out in section \ref{sec:Formalism}, where we consider perturbative solutions to the string field equation that represent RR background deformations, and outline the steps of extracting the string spectrum from the linearized string field fluctuations around the background.

In section \ref{sec:PlaneWave} we will consider type IIB strings in the pp-wave background supported by the self-dual RR 5-form flux as a basic example to illustrate our framework. We explicitly solve the string fields to leading nontrivial orders in the RR flux, and investigate linearized fluctuations that represent a family of maximally spinning string states. In particular, we compute corrections to their dispersion relations to quadratic order in the RR flux. In a rather intricate manner, our result is consistent with that of \cite{Metsaev:2001bj,Metsaev2002a,Berenstein:2002jq}, obtained by quantizing the Green-Schwarz action in the lightcone gauge.

In section \ref{sec:AdS3}, we consider type IIB strings in $AdS_3\times S^3\times M_4$ (where $M_4$ is $K3$ or $T^4$) in mixed three-form flux backgrounds, viewed as a deformation of the pure NSNS background by turning on RR flux. We will focus on pulsating string states due to the simple form of their vertex operators, and compute their mass corrections to quadratic order in the RR flux. The crux of the computation is a straightforward but tedious exercise of evaluating (and integrating) correlation functions of $SL(2,\mathbb{R})$ current algebra descendants, the details of which are given in the Appendices. 

Our explicit result for the mass corrections at finite $AdS$ radius is numerical. In the large radius and large oscillator number limit, our result agrees with that obtained from semiclassical quantization of a pulsating string in $AdS_3$ \cite{Hernandez:2018gcd}. In the Berenstein-Maldacena-Nastase (BMN) limit \cite{Berenstein:2002jq}, our result also agrees with the previously known string spectrum in the mixed flux pp-wave background obtained by lightcone quantization of the Green-Schwarz action.
We conclude and comment on future perspectives in section \ref{discsec}.

\section{Background deformation in NSR superstring field theory}\label{sec:Formalism}

\subsection{Closed NSR superstring field theory}
\label{csftsec}

A systematic treatment of background deformation in superstring perturbation theory is provided by the closed superstring field theory based on the NSR formalism, as formulated in \cite{Pius:2014iaa, Sen:2014pia, Sen:2014dqa, Sen:2015hha, deLacroix:2017lif}. In this paper we will only be concerned with classical string field theory. More precisely, we will work with the string field equation derived from the tree level 1PI off-shell amplitudes. Quantum corrections can be computed once the full 1PI effective action is considered, but they lie outside the scope of this work.

One defines the closed string field $\Psi$ as a state in the worldsheet CFT Hilbert space of picture number $-1$ in the NS sector and $-{1\over 2}$ in the R sector\footnote{In the notation of \cite{deLacroix:2017lif}, these are states in the space $\widehat{\cal H}_T$. But there are also states in the space $\widetilde{\cal H}_T$ which in particular involves picture number $-{3\over2}$ states in the R sector that play essential roles in the superstring field theory. However, the equations of motions for the fields in $\widetilde{\cal H}_T$ can be solved for, up to free field equations, once string fields in $\widehat{\cal H}_T$ are solved (see e.g. section 6.1 in \cite{deLacroix:2017lif}). Therefore, we will focus on the states in $\widehat{\cal H}_T$ in this work and assume that the corresponding solution for the fields in $\widetilde{\cal H}_T$ exists.}, that is subject to the constraints
\ie\label{levelmatchcons}
(b_0 - \widetilde b_0) |\Psi\rangle = (L_0 - \widetilde L_0) |\Psi\rangle = 0.
\fe
One further defines the operator ${\cal G}$ as the identity when acting on an NS sector state, or the zero mode of the PCO, ${\cal X}_0$ or $\widetilde{\cal X}_0$, when acting on an R sector state on the left or on the right, respectively. Our convention for the PCO ${\cal X}(z)$ is
\ie
{\cal X}(z) = Q_{\rm BRST} \cdot \xi(z) = -{1\over 2} e^\phi G^{\rm m} + c \partial \xi - {1\over 4} e^{2\phi} \partial\eta b - {1\over 4} \partial(e^{2\phi} \eta b),
\fe
where $G^{\rm m}$ is the supercurrent of the matter CFT. $\xi, \eta, \phi$ are the standard re-bosonization of the $\beta,\gamma$ ghost system,
\ie
\beta = e^{-\phi}\partial\xi,~\gamma = e^{\phi}\eta.
\fe

The string field equation takes the form
\ie\label{sfteom}
Q_B |\Psi\rangle + \sum_{n=2}^\infty {1\over n!} {\cal G} |[\Psi^{\otimes n}] \rangle = 0,
\fe
whereas gauge transformations are given by
\ie
\delta_\Lambda |\Psi\rangle = Q_B |\Lambda\rangle + \sum_{m=1}^\infty {1\over m!} {\cal G} |[\Psi^{\otimes m}\otimes \Lambda]\rangle.
\fe
Here $Q_B$ is the BRST operator in the worldsheet CFT. The bracket $[\Psi^{\otimes (n-1)}]$ is a multi-linear map on the string field defined through
\ie
\langle \Psi_1 | {c_0 - \widetilde c_0\over 2} | [\Psi_2\otimes\cdots \otimes\Psi_n]  \rangle = \{ \Psi_1\otimes\cdots \otimes\Psi_n\},
\fe
where $\{ \Psi_1\otimes\cdots \otimes\Psi_n\}$ is the 1PI part of the tree-level off-shell amplitude ${\cal A}[\Psi_1,\cdots,\Psi_n]$. That is, we subtract from ${\cal A}[\Psi_1,\cdots,\Psi_n]$ all 1-particle reducible contributions that correspond to off-shell amplitudes of sub-diagrams connected by free string field propagators. The off-shell amplitude ${\cal A}$, as defined in \cite{Pius:2014iaa, Sen:2014pia, deLacroix:2017lif}, takes the form
\ie\label{asint}
{\cal A}[\Psi_1,\cdots,\Psi_n] = \int_{{\cal S}_n} \Omega[\Psi_1,\cdots,\Psi_n],
\fe
where the integrand $\Omega$ and the integration domain ${\cal S}_n$ are defined as follows. In the holomorphic sector, suppose $n_{\rm NS}$ of the string fields among $\Psi_i$ are in the NS sector (of picture number $-1$), and $n_{\rm R}$ of them are in the R sector (of picture number $-{1\over 2}$). We then need $n_{\rm NS} + {1\over 2}n_{\rm R} - 2$ holomorphic PCO insertions. Similarly we need $\widetilde n_{\rm NS} + {1\over 2}\widetilde n_{\rm R} - 2$ antiholomorphic PCO insertions. Let ${\cal M}_{0,n}$ be the moduli space of the $n$-punctured Riemann sphere. ${\cal P}_{0,n}$ is the space of the $n$-punctured Riemann sphere together with the choice of a holomorphic coordinate system $w_i$ on a disc $D_i=\{|w_i|\leq 1\}$ containing each puncture, $i=1,\cdots, n$. $\widetilde{\cal P}_{0,n}$ is the space of the data that define ${\cal P}_{0,n}$ together with the choice of the locations of the PCOs, all of which are inserted outside of $\bigcup_{i=1}^n D_i$. ${\cal S}_n$ is a suitable subspace of $\widetilde{\cal P}_{0,n}$ that projects onto the entire ${\cal M}_{0,n}$ if we forget the data of $w_i$ and PCOs.

$\Omega$ is a degree $(2n-6)$ differential form on $\widetilde{\cal P}_{0,n}$ defined by a suitable correlation function of the string fields inserted at the punctures on the sphere, together with insertions of $b$ ghosts associated with the moduli, and insertions of PCOs. For instance, if we choose the coordinate $w_i$ on the disc $D_i$ to be related to the coordinate $z_i$ on the Riemann sphere by the $PSL(2,\mathbb{C})$ map
\ie\label{pslmap}
z = f_i(w_i) \equiv z_i + {q_i w_i\over 1+ s_i q_i w_i},
\fe
the form $\Omega$ can be written as\footnote{Compared to the convention of \cite{Pius:2014iaa, deLacroix:2017lif}, our string field is rescaled by a factor of $2\pi i$. }
\ie\label{omegadef}
\Omega[\Psi_1,\cdots,\Psi_n] &= 2\pi (-)^{n-1} \Bigg\langle \prod_{m=1}^{n_{\rm NS} + {1\over 2}n_{\rm R} - 2} \big({\cal X}(y_m) - \partial \xi(y_m) dy_m \big) \prod_{\widetilde m=1}^{\widetilde n_{\rm NS} + {1\over 2}\widetilde n_{\rm R} - 2} \big(\widetilde {\cal X}(\bar y_{\widetilde m}) - \bar\partial \widetilde \xi(\bar y_{\widetilde m}) d\bar y_{\widetilde m} \big)
\\
&~~~\times \exp\left[ \sum_{k=1}^{n-3} dt^k \sum_{i=1}^n \oint_{C_{z_i}} {dz\over 2\pi i} b(z) \left.{\partial z\over \partial t^k}\right|_{w_i} - \sum_{k=1}^{n-3} d\bar t^k \sum_{i=1}^{n} \oint_{C_{z_i}} {d\bar z\over 2\pi i} \widetilde b(\bar z) \left.{\partial \bar z\over \partial \bar t^k}\right|_{\bar w_i}  \right] 
\\
&~~~\times \left.  \prod_{i=1}^n e^{-s_i L_1 - \bar s_i \widetilde L_1} q_i^{L_0} \bar q_i^{\widetilde L_0}  \Psi_i(z_i,\bar z_i)  \Bigg\rangle\right|_{(n-3,n-3)} ,
\fe
where $C_{z_i}$ is a counterclockwise contour around $z_i$, and $\left.\cdots\right|_{(p,q)}$ means that we are keeping only the $(p,q)$-form component. The integration slice ${\cal S}_n$ is defined by a choice of $z_i$, $y_m$, $q_i$, $s_i$ as functions of the moduli $(t^k, \bar t^k)$ that parameterize ${\cal M}_{0,n}$. Note that the $b$ ghost insertion in (\ref{omegadef}) is multiplied by $\left.{\partial z\over \partial t^k}\right|_{w_i} = {\partial z_i\over \partial t^k} + o(z-z_i)$, where the $o(z-z_i)$ terms depend on the choice of $q_i$ and $s_i$ as functions of $t^k$. The result of the $b$ ghost contour integral is of the form $dz^i b_{-1} + dt^k \sum_{n\geq 0}h_{k,n}^i b_n$, for some coefficients $h_{k,n}^i$, acting on the string field inserted at $z_i$.

${\cal S}_n$ should be arranged such that the amplitude is completely symmetric in the $n$ string fields. To achieve this in practice it is convenient to average over different choices of ${\cal S}_n$. For instance, we can define the ``ordered off-shell amplitude" ${\cal A}^{\rm ord}[\Psi_1,\cdots,\Psi_n]$ by performing the integration (\ref{asint}) over a slice ${\cal S}_n^{\rm ord}$ defined through fixed $z_1=0$, $z_2=1$, $z_3=\infty$, with the moduli of ${\cal M}_{0,n}$ parameterized by $z_4, \cdots, z_n$, and an appropriate assignment of $y_m, q_i, s_i$ that respects permutation symmetry on $\Psi_4, \cdots, \Psi_n$. We then obtain the off-shell amplitude ${\cal A}$ by symmetrizing all string fields in ${\cal A}^{\rm ord}$. 

Note that deforming ${\cal S}_n$ amounts to a field redefinition of the string field.  In order for the amplitude to factorize consistently when an intermediate closed string state goes on-shell, one further demands that the choices of ${\cal S}_n$ for different $n$'s are compatible with respect to gluing together punctured spheres via plumbing fixture in the degeneration limits where the punctures cluster. Such choices are possible because the plumbing can be performed using $PSL(2,\mathbb{C})$ maps.

As an example, let us consider the 3-point ordered amplitude
\ie\label{aordthree}
{\cal A}^{\rm ord} [\Psi_1, \Psi_2, \Psi_3] = 2\pi \left\langle \prod_{m=1}^{n_{\rm NS} + {1\over 2}n_{\rm R} - 2} {\cal X}(y_m) \widetilde {\cal X}(\bar y_{\widetilde m}) \prod_{i=1}^3 e^{-s_i L_1 - \bar s_i \widetilde L_1} q_i^{L_0} \bar q_i^{\widetilde L_0}  \Psi_i(z_i,\bar z_i)  \right\rangle,
\fe
with the choice of ${\cal S}_3^{\rm ord}$, in this case a point in ${\widetilde{\cal P}}_{0,3}$, labeled by $z_i, y_m, q_i, s_i$. For simplicity let us ignore the PCO for the moment, and choose $z_1= 0$, $z_2=1$, $z_3=\infty$, with the transition maps (\ref{pslmap}) given by
\ie\label{pslthree}
f_1(w_1) = r w_1,~~~~ f_2(w_2) = 1 - r w_2,~~~~ f_3(w_3) = (r w_3)^{-1},
\fe
which amounts to $q_1=-q_2=q_3=r$, $s_1=s_2=s_3=0$. Note that our convention for the string field inserted at $z=\infty$ is such that $f_3$ is inverted.
The off-shell amplitude ${\cal A}[\Psi_1, \Psi_2, \Psi_3]$ is obtained from (\ref{aordthree}) by averaging over permutations on $\Psi_1, \Psi_2, \Psi_3$. Equivalently, we can average over 6 different choices of ${\cal S}_3^{\rm ord}$, related by the $PSL(2,\mathbb{C})$ maps permuting $0,1,\infty$, namely
\ie
z\mapsto z,~~1-z,~~{1\over z},~~ {1\over 1-z},~~ {z-1\over z},~~{z\over z-1}.
\fe
For instance, the map $z\mapsto 1-z$ which exchanges 0 and 1, together with $f_1\leftrightarrow f_2$, leaves $f_1, f_2$ invariant while changing $f_3$ to
\ie
f_3(w_3) = \left({-rw_3\over 1-rw_3}\right)^{-1}.
\fe
This corresponds to the choice $q_3=-r$, $s_3=1$. Thus, the insertion of the third string field takes the form $\left[ e^{-L_1-\widetilde L_1} r^{2L_0} \Psi_3 \right](\infty)$. Note that, if all three string fields $\Psi_i$ are $SL(2)\times SL(2)$ primaries, the ordered amplitude defined through the transition map (\ref{pslthree}) is the same as the symmetrized off-shell amplitude.

\subsection{RR deformation and Weyl invariance}
\label{rrweylsft}

An RR deformation of a closed string background corresponds to a solution $\Psi_0$ of the string field equation (\ref{sfteom}) of the form
\ie\label{psizero}
\Psi_0 = \sum_{n=2}^\infty \mu^n V_{{\rm NS}, n}^{-1,-1} + \sum_{n=1}^\infty \mu^n V_{{\rm R},n}^{-{1\over 2},-{1\over 2}},
\fe
where the superscripts indicate the picture numbers, and $\mu$ is the RR deformation parameter. The string field equation expanded to first order in $\mu$ implies that $V_{{\rm R}, 1}^{-{1\over 2},-{1\over 2}}$ is BRST closed, and thus takes the form of the usual on-shell RR vertex operator. It can be put in the form
\ie\label{psivv}
&V_{{\rm R}, 1}^{-{1\over 2},-{1\over 2}} = c\widetilde c e^{-{\phi\over 2} - {\widetilde\phi\over 2}} {\cal O}_{{\rm R}, 1},
\fe
where ${\cal O}_{{\rm R},1}$ is a matter CFT RR operator of weight $({5\over 8},{5\over 8})$.

The NSNS component of the string field equation at order $\mu^2$ is
\ie\label{quadsfteqn}
Q_B V_{{\rm NS}, 2}^{-1,-1} + {1\over 2} \left[\left(V_{{\rm R}, 1}^{-{1\over 2},-{1\over 2}}\right)^{\otimes 2}\right]  = 0.
\fe
If $\left[\left(V_{{\rm R}, 1}^{-{1\over 2},-{1\over 2}}\right)^{\otimes 2}\right]$ is not BRST-exact, it would present an obstruction to turning on the RR background. We will proceed by assuming that the string field solution does exist.
Let $\mathbb{P}$ be the projection operator onto states with $L_0 + \widetilde L_0 = 0$. A solution to (\ref{quadsfteqn}) can be written in the form
\ie\label{vwsep}
V_{{\rm NS}, 2}^{-1,-1} = W^{(2)}_{\rm NS} - {1\over 2}{1\over L_0 + \widetilde L_0} (b_0 + \widetilde b_0) (1-\mathbb{P}) \left[ \left(V_{{\rm R},1}^{-{1\over 2},-{1\over 2}} \right)^{\otimes 2}\right],
\fe
where $W^{(2)}_{\rm NS}$ is a state that obeys
\ie\label{wnseqn}
Q_B W^{(2)}_{\rm NS} = - {1\over 2} \mathbb{P} \left[ \left(V_{{\rm R},1}^{-{1\over 2},-{1\over 2}} \right)^{\otimes 2}\right].
\fe
In practice, we will be able to solve $W^{(2)}_{\rm NS}$ relatively easily, up to a BRST-closed term that must in principle be adjusted to allow for a solution at higher orders in $\mu$.

Typically, $V_{{\rm NS}, 2}^{-1,-1}$ takes the form
\ie\label{vnssketch}
&V_{{\rm NS}, 2}^{-1,-1} = c\widetilde c e^{-\phi - \widetilde\phi} {\cal O}_{{\rm NS}, 2} + \cdots,
\fe
where ${\cal O}_{{\rm NS},2}$ is a matter CFT operator, and $\cdots$ represents higher weight operators that generally involve other combinations of ghosts. To illustrate the point, let us contract (\ref{quadsfteqn}) with $\langle \Psi'| {c_0 - \widetilde c_0\over 2} = \langle \Psi'| {c_0 - \widetilde c_0\over 2} (c_0 b_0+\widetilde c_0 \widetilde b_0)$ for an arbitrary NSNS string field $\Psi'$ that obeys the Siegel gauge constraint $(b_0 + \widetilde b_0)\Psi' = 0$. Using $\{Q_B, b_0\} = L_0$, we have
\ie\label{sfteomcontr}
\langle \Psi'| c_0 \widetilde c_0 L_0 | V_{{\rm NS}, 2}^{-1,-1} \rangle + {1\over 2} \left\{ \Psi'\otimes \left(V_{{\rm R}, 1}^{-{1\over 2},-{1\over 2}}\right)^{\otimes 2} \right\} = 0.
\fe
Here $\{\cdots\}$ is the 1PI part of the off-shell amplitude introduced in section \ref{csftsec}, which in the three-point case coincides with the off-shell amplitude itself. The latter is particularly simple, as neither moduli nor PCO are involved. If we take $\Psi' = c\widetilde c e^{-\phi - \widetilde\phi} \Phi'$ for some matter CFT operator $\Phi'$ that is an $SL(2)\times SL(2)$ primary, then (\ref{sfteomcontr}) reduces to the following condition on ${\cal O}_{{\rm NS},2}$ at the level of matter CFT correlators,
\ie\label{simplaftem}
\left\langle \Phi' (\infty) \,\big(L_0-{1\over 2}\big) {\cal O}_{{\rm NS}, 2}(0)  \right\rangle + {\pi} \Big\langle  
r^{2L_0 - 1} \Phi'(\infty) \,{\cal O}_{{\rm R},1}(0) \, {\cal O}_{{\rm R},1}(1) \Big\rangle = 0,
\fe
where we have adopted the transition maps (\ref{pslthree}) in writing the 3-point off-shell amplitude.

From the point of view of deforming the worldsheet CFT along the lines of (\ref{rrdef}), one may identify the RR deformation operator with $\mu e^{-{\phi\over 2} - {\widetilde\phi\over 2}} {\cal O}_{{\rm R},1}$, and identify the NSNS deformation operator with the picture-raised version of $ \mu^2 e^{-\phi-\widetilde\phi} {\cal O}_{{\rm NS}, 2}$ (note however that ${\cal O}_{{\rm NS},2}$ is off-shell). (\ref{simplaftem}) amounts to the cancellation of the Weyl anomaly due to a pair of colliding RR deformation operators by that of the NSNS deformation operator, where the ``radius" parameter $r$ now plays the role of a UV cutoff on the worldsheet. In contrast, in the SFT approach the worldsheet CFT is never deformed, and changing the parameter $r$ amounts to a redefinition of the string field.

\subsection{String spectrum from linearized string field equation}

Expanding the string field $\Psi$ around a background solution $\Psi_0$, $\Psi = \Psi_0 + \widehat \Psi$, we can extract the string spectrum from the linearized equation in $\widehat \Psi$,
\ie\label{linearfluc}
Q_B |\widehat \Psi\rangle + \sum_{n=1}^\infty {1\over n!} {\cal G}|[\Psi_0^{\otimes n} \otimes\widehat\Psi]\rangle = 0,
\fe
modulo gauge redundancies of the form $\delta_\Lambda|\widehat\Psi\rangle = Q_B|\Lambda\rangle + \sum_{m=1}^\infty {1\over m!}{\cal G} |[\Psi_0^{\otimes m}\otimes\Lambda]\rangle$. 

Let us consider the RR background of the form (\ref{psizero}), (\ref{psivv}), and restrict our attention to bosonic fluctuations described by $\widehat\Psi$. We will split $\widehat\Psi$ into its NSNS component $\widehat\Psi_{\rm NS}$ and the RR component $\widehat\Psi_{\rm R}$, and expand each of them as a power series in $\mu$, 
\ie\label{nssffluc}
& \widehat\Psi_{\rm NS} = \sum_{n=0}^\infty \mu^{n} \widehat\Psi_{\rm NS}^{(n)},
~~~~ \widehat\Psi_{\rm R} = \sum_{n=0}^\infty \mu^{n} \widehat\Psi_{\rm R}^{(n)}. 
\fe 
At zeroth order in $\mu$, (\ref{nssffluc}) amounts to a linear combination of the BRST-closed NSNS state $\widehat\Psi_{\rm NS}^{(0)}$ and the RR state $\widehat\Psi_{\rm R}^{(0)}$ in the original undeformed background. The deformed solution in particular encodes corrections to the dispersion relation, or ``mass", of the string state in question.

One should be cautious that, while it is typically possible to arrange $\widehat\Psi$ to be a delta-function normalizable state with respect to spacetime momenta, the series expansion of $\widehat\Psi$ to a given order in $\mu$ are not quite delta-function normalizable.\footnote{While the normalizability issue of the $\mu$-expansion of the string field does not pose an essential difficulty for us, it may be circumvented by considering an alternative expansion scheme, where the zeroth order string field is taken off-shell and involves the corrected spacetime momentum. We thank Ashoke Sen for discussions on this point.} For instance, if $\widehat\Psi^{(0)}_{\rm NS/R}$ contains $e^{ik\cdot X}$ for some spacetime momentum $k^\mu$, $\widehat\Psi$ may involve $e^{ik'\cdot X}$ for a shifted spacetime momentum $k'^\mu = k^\mu + \delta k^\mu$ that obeys a corrected dispersion relation. $\widehat\Psi_{\rm NS/R}^{(1)}$ would involve states like $\delta k\cdot X e^{ik\cdot X}$ whose inner products are not delta-functions in $k$ but rather derivatives of delta-functions. Acting on states whose inner products are defined in this distributional sense, $L_0$ is not diagonalizable (as is standard in logarithmic CFTs). Instead, we can work with a basis of states on which $L_0$ acts in its Jordan normal form. The projector $\mathbb{P}$ onto $L_0=0$ states defined in section \ref{rrweylsft} will now be generalized to the projection operator onto the invariant subspace on which $L_0$ acts nilpotently.

\subsubsection{First order in $\mu$ and mixing between NSNS and RR states}

Expanding the equation (\ref{linearfluc}) to first order in $\mu$ gives
\ie\label{qbfirstp}
 Q_B \widehat\Psi_{\rm R}^{(1)} +  {\cal X}_0 \widetilde{\cal X}_0\left[ V_{{\rm R}, 1}^{-{1\over 2},-{1\over 2}} \otimes \widehat\Psi_{\rm NS}^{(0)} \right] =& 0,
\\
 Q_B \widehat\Psi_{\rm NS}^{(1)} +  \left[ V_{{\rm R}, 1}^{-{1\over 2},-{1\over 2}} \otimes \widehat\Psi_{\rm R}^{(0)} \right] =& 0.
\fe
We can write the solution for the first order string field in the form
\ie\label{psirrsol}
& \widehat\Psi_{\rm R}^{(1)} =  Y_{\rm R}^{(1)} - {1\over L_0 + \widetilde L_0} (b_0 + \widetilde b_0) (1-\mathbb{P}) {\cal X}_0 \widetilde{\cal X}_0  \left[ V_{{\rm R}, 1}^{-{1\over 2},-{1\over 2}} \otimes \widehat\Psi_{\rm NS}^{(0)} \right],
\\
&  \widehat\Psi_{\rm NS}^{(1)} =  Y_{\rm NS}^{(1)} - {1\over L_0 + \widetilde L_0} (b_0 + \widetilde b_0) (1-\mathbb{P}) \left[ V_{{\rm R}, 1}^{-{1\over 2},-{1\over 2}} \otimes \widehat\Psi_{\rm R}^{(0)} \right],
\fe
where $Y_{\rm R}^{(1)}$, $Y_{\rm NS}^{(1)}$ are $\mathbb{P}$-invariant states that obey
\ie\label{yreqn}
& Q_B Y_{\rm R}^{(1)} = - \mathbb{P} {\cal X}_0 \widetilde{\cal X}_0  \left[ V_{{\rm R}, 1}^{-{1\over 2},-{1\over 2}}\otimes \widehat\Psi_{\rm NS}^{(0)} \right],
\\
& Q_B Y_{\rm NS}^{(1)} = - \mathbb{P} \left[ V_{{\rm R}, 1}^{-{1\over 2},-{1\over 2}}\otimes \widehat\Psi_{\rm R}^{(0)} \right].
\fe
A priori, the solution for $Y_{\rm NS/R}^{(1)}$ exists only if the string field brackets appearing on the RHS (\ref{yreqn}) are BRST-exact. However, as already remarked, $Y_{\rm NS/R}^{(1)}$ are not necessarily delta-function normalizable states, which allows for the possibility of trivializing a larger class of BRST-closed states.

We will focus on solutions that represent string states of well-defined dispersion relation in the deformed background. Suppose the $\mu^0$ order string field takes the form
\ie{}
& \widehat\Psi_{\rm NS}^{(0)} = c\widetilde c \, e^{-\phi - \widetilde\phi} \Phi_{\rm NS}^m(k),
~~~~ \widehat\Psi_{\rm R}^{(0)} = c\widetilde c \, e^{-{\phi\over 2} - {\widetilde\phi\over 2}} \Phi_{\rm R}^m(k),
\fe
where $\Phi_{\rm NS}^{\rm m}(k)$ is a matter superconformal primary, and $\Phi^{\rm m}_{\rm R}$ is an RR state in the matter CFT that is annihilated by $G_r^{\rm m}$ and $\widetilde G_r^{\rm m}$ for $r\geq 0$. The parameter $k$ represents the spacetime momentum, or the analogous quantum numbers if our starting point is a nontrivial NSNS background. We will assume a continuation of $\Phi_{\rm NS/R}^{\rm m}(k)$ to off-shell values of $k$ such that $\Phi_{\rm NS/R}^{\rm m}(k)$ are still annihilated by $L^{\rm m}_{n> 0}$ and $G^{\rm m}_{r> 0}$ in the matter CFT.

Up to order $\mu$, we will demand that $\widehat\Psi_{\rm NS/R}^{(0)} + \mu Y_{\rm NS/R}^{(1)}$ are given by analogous expressions with $\Phi_{\rm NS/R}^m(k)$ replaced by $\Phi_{\rm NS/R}^m(k+\delta k)$, where $k+\delta k$ obeys a deformed dispersion relation, up to states that involve other ghost structures as well as matter super-Virasoro descendants. In other words, 
\ie\label{deomyy}
\mu Y_{\rm NS}^{(1)} = c\widetilde c \, e^{-\phi - \widetilde\phi} \delta k{\partial\over \partial k} \Phi_{\rm NS}^m(k) + \cdots,~~~~
\mu Y_{\rm R}^{(1)} = c\widetilde c \, e^{-{\phi\over 2} - {\widetilde\phi\over 2}} \delta k{\partial\over \partial k} \Phi_{\rm R}^m(k) + \cdots,
\fe
where $\cdots$ represent orthogonal states that either involve other ghost structures or matter SCFT descendants. 
Comparing with (\ref{yreqn}), we need
\ie\label{phinsrare}
&  (\partial c +\bar\partial\widetilde c) c\widetilde c \, e^{-{\phi\over 2} - {\widetilde\phi\over 2}} \big(L_0 - {5\over 8}\big) \delta k{\partial\over \partial k} \Phi_{\rm R}^m(k) + \cdots = - \mu\, \mathbb{P} {\cal X}_0 \widetilde{\cal X}_0  \left[ V_{{\rm R}, 1}^{-{1\over 2},-{1\over 2}}\otimes c\widetilde c \, e^{-\phi - \widetilde\phi} \Phi_{\rm NS}^m(k) \right] ,
\\
&   (\partial c +\bar\partial\widetilde c) c\widetilde c \,e^{-\phi - \widetilde\phi}  \big(L_0 - {1\over 2}\big) \delta k{\partial\over \partial k} \Phi_{\rm NS}^m(k) + \cdots = - \mu\,\mathbb{P} \left[ V_{{\rm R}, 1}^{-{1\over 2},-{1\over 2}}\otimes c\widetilde c \, e^{-{\phi\over 2} - {\widetilde\phi\over 2}} \Phi_{\rm R}^m(k) \right].
\fe
(\ref{phinsrare}) will determine the RR deformed dispersion relation as well as the mixing of NSNS and RR string states to first order in $\mu$.

\subsubsection{Second order in $\mu$}

Let us proceed to analyze the order $\mu^2$ string field equation for the linearized fluctuations,
\ie\label{secqbasft}
 Q_B \widehat\Psi_{\rm NS}^{(2)} + \left[ V_{{\rm R}, 1}^{-{1\over 2},-{1\over 2}} \otimes \widehat\Psi_{\rm R}^{(1)} +V_{{\rm NS}, 2}^{-1,-1} \otimes \widehat\Psi_{\rm NS}^{(0)} + {1\over 2} \left( V_{{\rm R}, 1}^{-{1\over 2},-{1\over 2}}\right)^{\otimes 2} \otimes \widehat\Psi_{\rm NS}^{(0)} \right] =& 0,
\\
 Q_B \widehat\Psi_{\rm R}^{(2)} + {\cal X}_0\widetilde{\cal X}_0\left[ V_{{\rm R}, 1}^{-{1\over 2},-{1\over 2}} \otimes \widehat\Psi_{\rm NS}^{(1)} +V_{{\rm NS}, 2}^{-1,-1} \otimes \widehat\Psi_{\rm R}^{(0)} + {1\over 2} \left( V_{{\rm R}, 1}^{-{1\over 2},-{1\over 2}}\right)^{\otimes 2} \otimes \widehat\Psi_{\rm R}^{(0)} \right] =& 0,
\fe
whose solution takes the form
\ie{}
& \widehat\Psi_{\rm NS}^{(2)} = Y_{\rm NS}^{(2)} - {1\over L_0 + \widetilde L_0}(b_0+\widetilde b_0) (1-\mathbb{P}) \left[ V_{{\rm R}, 1}^{-{1\over 2},-{1\over 2}} \otimes \widehat\Psi_{\rm R}^{(1)} +V_{{\rm NS}, 2}^{-1,-1} \otimes \widehat\Psi_{\rm NS}^{(0)} + {1\over 2} \left( V_{{\rm R}, 1}^{-{1\over 2},-{1\over 2}}\right)^{\otimes 2} \otimes \widehat\Psi_{\rm NS}^{(0)} \right],
\\
& \widehat\Psi_{\rm R}^{(2)} = Y_{\rm R}^{(2)} - {1\over L_0 + \widetilde L_0}(b_0+\widetilde b_0) (1-\mathbb{P}) {\cal X}_0\widetilde{\cal X}_0 \left[ V_{{\rm R}, 1}^{-{1\over 2},-{1\over 2}} \otimes \widehat\Psi_{\rm NS}^{(1)} +V_{{\rm NS}, 2}^{-1,-1} \otimes \widehat\Psi_{\rm R}^{(0)} + {1\over 2} \left( V_{{\rm R}, 1}^{-{1\over 2},-{1\over 2}}\right)^{\otimes 2} \otimes \widehat\Psi_{\rm R}^{(0)} \right],
\fe
where $Y_{\rm NS/R}^{(2)}$ obeys the equation restricted to the $\mathbb{P}$-invariant subspace,
\ie\label{qbynseqn}
 Q_B Y_{\rm NS}^{(2)} + \mathbb{P}\left[ V_{{\rm R}, 1}^{-{1\over 2},-{1\over 2}} \otimes \widehat\Psi_{\rm R}^{(1)} +V_{{\rm NS}, 2}^{-1,-1} \otimes \widehat\Psi_{\rm NS}^{(0)} + {1\over 2} \left( V_{{\rm R}, 1}^{-{1\over 2},-{1\over 2}}\right)^{\otimes 2} \otimes \widehat\Psi_{\rm NS}^{(0)} \right] =& 0,
\\
 Q_B Y_{\rm R}^{(2)} + \mathbb{P}{\cal X}_0\widetilde{\cal X}_0\left[ V_{{\rm R}, 1}^{-{1\over 2},-{1\over 2}} \otimes \widehat\Psi_{\rm NS}^{(1)} +V_{{\rm NS}, 2}^{-1,-1} \otimes \widehat\Psi_{\rm R}^{(0)} + {1\over 2} \left( V_{{\rm R}, 1}^{-{1\over 2},-{1\over 2}}\right)^{\otimes 2} \otimes \widehat\Psi_{\rm R}^{(0)} \right] =& 0.
\fe
Analogously to (\ref{deomyy}), in order to have a well-defined dispersion relation, we need
\ie\label{masslzero}
& \widehat\Psi_{\rm NS}^{(0)} + \mu Y_{\rm NS}^{(1)} + \mu^2 Y_{\rm NS}^{(2)} = c\widetilde c \, e^{-\phi - \widetilde\phi}  \Phi_{\rm NS}^m(k+\delta k) +\cdots,
\\
&\widehat\Psi_{\rm R}^{(0)} + \mu Y_{\rm R}^{(1)} + \mu^2 Y_{\rm R}^{(2)} = c\widetilde c \, e^{-{\phi\over 2} - {\widetilde\phi\over 2}}  \Phi_{\rm R}^m(k+\delta k)  + \cdots.
\fe
Assuming this is the case, we can determine the order $\mu^2$ correction to the dispersion relation from $L_0 Y_{\rm NS/R}^{(2)}$. To compute the latter, we can contract (\ref{qbynseqn}) with $\langle \Psi'| {c_0 - \widetilde c_0\over 2} = \langle \Psi'| {c_0 - \widetilde c_0\over 2} (c_0 b_0+\widetilde c_0 \widetilde b_0)$ for some $\Psi'$ obeying the Siegel gauge constraint. Using (\ref{vwsep}), we have
\ie\label{simpfinaleqn}
& \left\langle \Psi'\Big| c_0 \widetilde c_0 L_0 Y_{\rm NS}^{(2)} \right\rangle +  \left\{ \mathbb{P}\Psi'\otimes \left( W^{(2)}_{\rm NS}\otimes \widehat\Psi_{\rm NS}^{(0)} + V_{{\rm R}, 1}^{-{1\over 2},-{1\over 2}} \otimes Y_{\rm R}^{(1)} \right) \right\} + {1\over 2} {\cal A}' \left[ \mathbb{P}\Psi'\otimes \left( V_{{\rm R}, 1}^{-{1\over 2},-{1\over 2}} \right)^{\otimes 2} \otimes \widehat\Psi_{\rm NS}^{(0)} \right] = 0,
\fe
where we have defined (with $\Psi''=\mathbb{P}\Psi'$)
\ie\label{modamp}
 &{\cal A}' \left[ \Psi''\otimes \left( V_{{\rm R}, 1}^{-{1\over 2},-{1\over 2}} \right)^{\otimes 2} \otimes \widehat\Psi_{\rm NS}^{(0)} \right]\\
  =& \left\{ \Psi''\otimes \left( V_{{\rm R}, 1}^{-{1\over 2},-{1\over 2}} \right)^{\otimes 2} \otimes \widehat\Psi_{\rm NS}^{(0)} \right\} - \left\{ \Psi''\otimes \widehat\Psi_{\rm NS}^{(0)} \otimes  {1\over L_0 + \widetilde L_0} (b_0 + \widetilde b_0) (1-\mathbb{P}) \left[ \left(V_{{\rm R},1}^{-{1\over 2},-{1\over 2}} \right)^{\otimes 2}\right]  \right\}
\\
& - 2 \left\{ \Psi''\otimes V_{{\rm R}, 1}^{-{1\over 2},-{1\over 2}} \otimes {1\over L_0 + \widetilde L_0} (b_0 + \widetilde b_0) (1-\mathbb{P}) {\cal X}_0 \widetilde{\cal X}_0  \left[ V_{{\rm R}, 1}^{-{1\over 2},-{1\over 2}} \otimes \widehat\Psi_{\rm NS}^{(0)} \right] \right\} .
\fe 
We will refer to ${\cal A}'$ as a {\it modified amplitude} of string fields, in the following sense. If we omit the projector $\mathbb{P}$ on the RHS of (\ref{modamp}), it would give ${1\over 2}$ times the off-shell tree level 4-point amplitude ${\cal A}$ of the string fields $\Psi''$, $V_{{\rm R},1}$, $V_{{\rm R},1}$, and $\widehat\Psi_{\rm NS}^{(0)}$. ${\cal A}'$ differs from ${\cal A}$ in that, when the latter diverges as an intermediate string state going on-shell, a pole contribution is subtracted off so that ${\cal A}'$ is finite. From the point of view of the spacetime S-matrix, this would be a contrived definition. The amplitude considered here involves the insertion of background string fields whose momenta may not be adjustable, and the subtraction is useful in organizing practical computations.

We can compute the 4-point off-shell amplitude ${\cal A}$ by averaging the ordered amplitude ${\cal A}^{\rm ord}$ over permutations on the four string fields. For instance, one of the ordered amplitudes is\footnote{We adopt the measure convention $dz^2=|dz\wedge d\bar z|$ as in \cite{polchinski1998string}.}
\ie\label{offshellam}
& {\cal A}^{\rm ord} \left[ \widehat\Psi_{\rm NS}^{(0)}\otimes V_{{\rm R}, 1}^{-{1\over 2},-{1\over 2}} \otimes \Psi''\otimes  V_{{\rm R}, 1}^{-{1\over 2},-{1\over 2}} \right]\\
=& -2\pi \int d^2z \Bigg\langle \left[ e^{-sL_1-s\widetilde L_1}r^{2L_0}\Psi'' \right](\infty)\, V_{{\rm R},1}^{-{1\over 2},-{1\over 2}}(1) \widehat\Psi_{\rm NS}^{(0)}(0) \bigg\{ \big[{\cal X}(y) b_{-1}-\partial_z \xi(y) \big] \big[\widetilde{\cal X}(\bar y) \widetilde b_{-1}-\bar\partial_{\bar z} \widetilde\xi(\bar y) \big]\\
& -{\cal X}(y) \partial_z \widetilde \xi(\bar y) \widetilde b_{-1} + \widetilde{\cal X}(\bar y) \bar\partial_{\bar z} \xi(y) b_{-1} - \bar\partial_{\bar z} \xi(y) \partial_z \widetilde \xi(\bar y) \bigg\} V_{{\rm R},1}^{-{1\over 2},-{1\over 2}}(z,\bar z)
\Bigg\rangle,
\fe
where $s$ and $r=|q|$ are parameters of the transition map for the coordinate system on the disc containing $\Psi''(\infty)$, which are generally functions of $(z,\bar z)$ to ensure the correct factorization limits. In writing (\ref{offshellam}) we have used the fact that $V_{{\rm R},1}$ and $\widehat\Psi_{\rm NS}^{(0)}$ are on-shell.
A priori, the location of the PCOs $y=y(z, \bar z)$ as specified by the integration slice ${\cal S}\subset \widetilde{P}_{0,4}$ should be arranged such that when $z$ approaches either 0, 1, or $\infty$, $y$ is away from $z$. If $\Psi''$ is on-shell, then the form $\Omega$ (\ref{omegadef}) is closed, and one can deform $\cal S$ by moving the PCOs to $z$, and convert $V_{{\rm R},1}^{-{1\over 2}, -{1\over 2}}$ to a string field in the $({1\over 2},{1\over 2})$ picture, up to a possible boundary term of the moduli integral that can typically be shown to vanish. When $\Psi''$ is off-shell, deforming $\cal S$ to $\cal S'$ ``vertically" by moving the PCOs changes the amplitude ${\cal A}$ by the integral of $d\Omega$ along a chain in $\widetilde{P}_{0,4}$ bounded by ${\cal S}'-{\cal S}$. The latter in particular involves the ``vertical" integration of $d\Omega$, with the latter expressed as a correlator involving $Q_B\Psi''$ and explicit insertions of $\partial \xi$ or $\bar\partial \widetilde\xi$. In simple situations where $Q_B\Psi''$ does not involve $\eta$ or $\widetilde\eta$, such vertical integration gives a vanishing result.

The corresponding ordered modified amplitude ${\cal A}'$ differs from (\ref{offshellam}) by subtracting off possible log divergences in the integration near $z=0, 1, \infty$. Note that the finite part of the subtraction is generally not invariant under field redefinition of the string field.

\section{Type IIB strings in pp-wave with RR 5-form flux}\label{sec:PlaneWave}

As the first example, we consider the simplest nontrivial solution of type IIB supergravity with RR flux, namely the pp-wave solution supported by the self-dual 5-form flux \cite{Metsaev:2001bj,Metsaev2002a,Berenstein:2002jq},
\ie\label{ppmetric}
& ds^2 = -2 dx^+ dx^- - \mu^2 \sum_{i=1}^8 z_i^2\, (dx^+)^2 + \sum_{i=1}^8 dz_i^2,
\\
& F^{\rm RR}_5 = \mu\,  dx^+\wedge (dz_1\wedge dz_2 \wedge dz_3\wedge dz_4 + dz_5\wedge dz_6 \wedge dz_7\wedge dz_8). 
\fe
We will first analyze the string field solution corresponding to the background (\ref{ppmetric}) to quadratic order in $\mu$, and will then study the spectrum of a family of maximally spinning strings from the linearized fluctuations of the string field.

\subsection{The background string field}

Viewed as a deformation of Minkowskian spacetime, the order $\mu$ string field takes the form of the vertex operator corresponding to the RR 5-form field strength. In the notation of (\ref{psizero}) and (\ref{psivv}), we have
\ie\label{vnsrsons}
& {\cal O}_{{\rm R},1} = N_1 S \Gamma^{+1234} (1+\Gamma) \widetilde S.
\fe
Here $S_\A$ and $\widetilde S_\A$ are the left and right spin fields, where we follow the conventions of \cite{polchinski1998string} and have included both chiral and anti-chiral components. The overall normalization constant $N_1$ will be determined shortly.

At order $\mu^2$, the NSNS string field takes the form (\ref{vwsep}), with the equation for $W^{(2)}_{\rm NS}$ (\ref{wnseqn}) now explicitly written as
\ie
Q_B W^{(2)}_{\rm NS} = 64\pi N_1^2 (\partial c + \bar\partial \widetilde c) c\widetilde c e^{-\phi - \widetilde\phi} \psi^+ \widetilde\psi^+.
\fe
This can be solved with
\ie\label{wnssimpsol}
W^{(2)}_{\rm NS} = - {16\pi\over \A'} N_1^2 c\widetilde c e^{-\phi - \widetilde\phi} \psi^+ \widetilde\psi^+ Z^2 ,
\fe
up to a $Q_B$-closed term, where $Z^2 = \sum_{i=1}^8 Z_i Z_i$ is understood to be normal ordered. Note that $Z^2$ is not an eigenstate of the dilatation operator, as $L_0 Z^2 = -4 \A'$. Furthermore, we are free to shift $Z^2$ in (\ref{wnssimpsol}) by a constant. In the supergravity description this amounts to shifting the metric by a constant times $\mu^2 (dx^+)^2$, which can be absorbed into a redefinition of the coordinate $x^-$.

It is useful to consider the picture-raised string field components,
\ie\label{vrnsaa}
V_{{\rm R},1}^{{1\over 2},{1\over 2}} = c\widetilde c e^{{\phi\over 2} + {\widetilde\phi\over 2}} {\cal O}_{{\rm R},1}^\uparrow,~~~~ V_{{\rm NS},2}^{0,0} = c\widetilde c {\cal O}^{\uparrow}_{{\rm NS}, n}+\cdots,
\fe
where  ${\cal O}^{\uparrow}_{{\rm R}, 1}$ ad ${\cal O}^{\uparrow}_{{\rm NS}, 2}$ are matter CFT operators, given by
\ie{}
&{\cal O}_{{\rm R},1}^\uparrow = {N_1\over 4} \,  \partial X^\mu \bar\partial X^\nu S\Gamma_\mu \Gamma^{+1234} (1+\Gamma)\Gamma_\nu \widetilde S,
\\
& {\cal O}_{{\rm NS},2}^\uparrow = -{4\pi N_1^2 \over \A'} \left( {2\over \A'} Z^2 \partial X^+ \bar\partial X^+ - 2 Z_i \psi_i \psi^+ \bar\partial X^+ - 2 Z_i \partial X^+ \widetilde\psi_i \widetilde\psi^+ + \A' \psi_i \psi^+ \widetilde\psi_i \widetilde\psi^+ \right) ,
\fe
whereas $\cdots$ in (\ref{vrnsaa}) stands for higher weight operators, which also involve more general combinations of ghost fields. From the point of view of deforming the worldsheet CFT along the lines of (\ref{rrdef}), ${\cal O}_{{\rm NS},2}^\uparrow$ can be identified with the worldsheet Lagrangian deformation. This also suggests that no further BRST-closed terms are needed in (\ref{wnssimpsol}). Comparison with the NSNS part of the supergravity solution (\ref{ppmetric}) then fixes the normalization constant
\ie
N_1 = {\sqrt{\A'}\over 4\pi}.
\fe
Since ${\cal O}_{{\rm NS},2}^\uparrow$ is not marginal, a deformation of the worldsheet CFT by ${\cal V}_{\rm NS}^{0,0} = \mu^2 {\cal O}_{{\rm NS},2}^\uparrow$ would lead to a  Weyl anomaly of the form $T^a{}_a = -16 \mu^2 \partial X^+ \bar\partial X^+$, consistent with the Ricci cuvature of (\ref{ppmetric}) having the only nonzero component $R_{++}=8\mu^2$. This is cancelled against the Weyl anomaly due to the logarithmic divergence that arises when a pair of RR deformation operators ${\cal V}_{\rm R}^{-{1\over 2},-{1\over 2}} = \mu e^{-{\phi\over 2} - {\widetilde\phi\over 2}} {\cal O}_{{\rm R},1}$ and ${\cal V}_{\rm R}^{{1\over 2},{1\over 2}} = \mu e^{{\phi\over 2} + {\widetilde\phi\over 2}} {\cal O}_{{\rm R},1}^\uparrow$ collide,
\ie\label{rrlog}
{1\over 2} \int d^2z d^2z' \, {\cal V}_{\rm R}^{-{1\over 2},-{1\over 2}}(z,\bar z) {\cal V}_{\rm R}^{{1\over 2},{1\over 2}}(z',\bar z')
&\sim -{\mu^2\over\pi^2} \int {d^2z d^2z'\over |z-z'|^2} \partial X^+(z') \bar\partial X^+(\bar z')
\\
&\sim{4\mu^2\over \pi}\ln\delta\int d^2z\, \partial X^+(z) \bar\partial X^+(\bar z),
\fe
where $\delta$ is a short-distance cutoff.

\subsection{The maximally spinning string states}
\label{rrcontsec}

Let us consider a family of NSNS string states on the leading Regge trajectory in Minkowskian spacetime, described by the vertex operator or string field
\ie\label{NSNSregge}
& \widehat\Psi_{\rm NS}^{(0)} =  c\widetilde c e^{-\phi-\widetilde\phi} e^{i(k_+ X^+ + k_- X^-)} 
F(Z_3,\cdots, Z_8) {\cal O}_{\rm osc} ,
\\
&{\cal O}_{\rm osc} = {1\over 2\A'^n n!}(\psi_1+i\psi_2)(\widetilde \psi_1+i\widetilde\psi_2) (\partial Z_1+i \partial Z_2)^n (\bar\partial Z_1+i \bar\partial Z_2)^n,
\fe
where the operator product is understood to be normal ordered. For simplicity, we have taken the wave function $F$ to be independent of $Z_1, Z_2$. This is such that (\ref{NSNSregge}) is BRST-closed, provided that 
\ie\label{fonsehll}
\left( 2k^+k^- + \partial_{Z_i} \partial_{Z_i} - {4n\over \A'} \right) F = 0.
\fe

In the pp-wave background, the order $\mu$ string field equation (\ref{yreqn}) for $Y_{\rm R}^{(1)}$ is very simple, as one can verify that
\ie\label{yrzeroa}
\mathbb{P}\left[ V_{{\rm R}, 1}^{-{1\over 2},-{1\over 2}}\otimes \widehat\Psi_{\rm NS}^{(0)} \right]=0
\fe
for $\widehat\Psi_{\rm NS}^{(0)}$ of the form (\ref{NSNSregge}). We will proceed by setting $Y_{\rm R}^{(1)}$ to zero. 

At order $\mu^2$, we will compute $Y_{\rm NS}^{(2)}$ from (\ref{simpfinaleqn}). First, we need to compute the projected string field bracket $\mathbb{P}\left[ W^{(2)}_{\rm NS} \otimes \widehat\Psi_{\rm NS}^{(0)} \right]$, which by definition can be extracted from the 3-point amplitude $\left\{ {\mathbb P}\Psi'\otimes W^{(2)}_{\rm NS} \otimes \widehat\Psi_{\rm NS}^{(0)} \right\}$ with a general off-shell string field $\Psi'$. The 3-point off-shell amplitude can be obtained by averaging the ordered amplitude ${\cal A}^{\rm ord}$ over permutations of the three string fields, as described in section \ref{csftsec}. Using the fact that $W^{(2)}_{\rm NS}$ and $\widehat\Psi_{\rm NS}^{(0)}$ are $SL(2)\times SL(2)$ primaries, we can write
\ie\label{awpsithree}
{\cal A}\left[  {\mathbb P}\Psi' \otimes W^{(2)}_{\rm NS} \otimes\widehat\Psi_{\rm NS}^{(0)} \right] 
=& 2\pi  \left\langle \widehat\Psi_{\rm NS}^{(0)}(0) \, r^{2L_0}W^{(2)}_{NS}(1) \, {1 + e^{-L_1-\widetilde L_1} \over 2} r^{2L_0}{\mathbb P}\Psi'(\infty) {\cal X}(y)\widetilde{\cal X}(\bar y) \right\rangle
\\
=& 2\pi \left\langle {1\over 2}\left[ \widehat\Psi_{\rm NS}^{(0)}(0) \, r^{2L_0}W^{(2)}_{NS}(1) + \widehat\Psi_{\rm NS}^{(0)}(-1) r^{2L_0}W^{(2)}_{NS}(0) \right] r^{2L_0}{\mathbb P}\Psi'(\infty) {\cal X}(y)\widetilde{\cal X}(\bar y) \right\rangle ,
\fe
where it is understood that the result should be averaged over a set of locations $(y, \bar y)$ of the PCO to ensure symmetry with respect to the three string field insertions. 

If $\Psi'$ takes the form $c\widetilde c e^{-\phi - \widetilde\phi}\Phi'$ for some matter superconformal primary $\Phi'$ (not necessarily marginal), so that $Q_B \Psi'$ does not involve $\eta$, then we would be free to move the PCOs in (\ref{awpsithree}). From the 3-point off-shell amplitude with such $\Psi'$, we can determine the projected string field bracket to be of the form
\ie\label{pwnsm}
\mathbb{P}\left[ W^{(2)}_{\rm NS} \otimes \widehat\Psi_{\rm NS}^{(0)} \right]
&=  {\A' \over 4} k_-^2  (\partial c + \bar\partial\widetilde c)c\widetilde c e^{-\phi - \widetilde\phi}  \left( 2\A' n +Z^2 - 8\A' \ln r \right) e^{i(k_+X^++k_-X^-)} F(Z) {\cal O}_{\rm osc} + \cdots ,
\fe
where 
$\cdots$ stands for states that involve either other ghost structures or matter superconformal descendants.

Next, we will examine the modified amplitude ${\cal A}'\left[ {\mathbb P}{\Psi}' \otimes \left( V_{{\rm R}, 1}^{-{1\over 2},-{1\over 2}} \right)^{\otimes 2} \otimes \widehat\Psi_{\rm NS}^{(0)} \right]$. For our purpose it will suffice to consider $\Psi' = c\widetilde c e^{-\phi-\widetilde\phi} \left[e^{-i(k_+'X^++k'_-X^-)} \overline{F'}(Z_1,\cdots, Z_8) \overline{\cal O}_{\rm osc}\right]_{\rm prim}$, where the subscript ``prim" stands for projection onto the matter superconformal primary component. Note that ${\mathbb P}\Psi'=\Psi'$, even though such $\Psi'$ is not necessarily BRST-closed. In this case, the corresponding 4-point off-shell amplitude of the string fields evaluates to
\ie\label{ampgs}
& {\cal A}\left[ \Psi' \otimes \left( V_{{\rm R}, 1}^{-{1\over 2},-{1\over 2}} \right)^{\otimes 2} \otimes \widehat\Psi_{\rm NS}^{(0)} \right]
\\
&= {\A'^2 k_-^2\over 2^9 \pi} \langle k', F'|k,F\rangle \int d^2z { |1-z|^{1\over 2}\over  |z|} \Big\langle (S(1) \Gamma_+\Gamma^{+1234}(1+\Gamma)\Gamma_+\widetilde S(1))(S(z) \Gamma^{+1234}(1+\Gamma)\widetilde S(\bar z)) 
\\
&~~~ \times (\psi_1+i\psi_2)(\widetilde\psi_1+i\widetilde\psi_2)(0) (\psi_1-i\psi_2)(\widetilde\psi_1-i\widetilde\psi_2)(\infty)  \Big\rangle
\\
&=-{\A'^2k_-^2\over 2\pi} \langle k',F'|k,F\rangle \int{d^2 z\over |z|^2}\left( 2\left|{z\over 1-z}\right|^2+{z\over 1-z}+{\bar z\over 1-\bar z}\right) 
\\
&= {4\A'^2 k_-^2}  \langle k', F'| k,F\rangle (\ln\D + \lambda),
\fe
where $\langle k',F'| k,F\rangle =(2\pi)^2 \delta^2(k-k') \int d^8Z\, \overline{F'(Z)} F(Z)$. In deriving the first equality above, we have moved the PCOs to convert one of the RR background string fields to $({1\over 2},{1\over 2})$ picture, and reduced the correlator in the moduli integrand to one in the free fermion CFT. The latter is then evaluated using $\langle S_\A(1)S_\B(z)\psi^\mu(0)\psi^\nu(\infty)\rangle=z^{-{1\over2}} (1-z)^{-{1\over4} }\left[{z \over 1-z}\eta^{\mu\nu}C_{\A\B}+{1\over2}\left(\Gamma^\mu\Gamma^\nu\right)_{\A\B} \right]$, where $C$ is the charge conjugation matrix on $so(1,9)$ spinors, and indices on gamma matrices are raised/lowered with $C$.

In the last step of (\ref{ampgs}), the $z$-integral is regularized by restricting to $|z-1|>\delta$ for a short-distance cutoff $\delta$, and $\lambda$ is a finite constant. Note the absence of log divergence near $z=0$ or $\infty$, which is consistent with (\ref{yrzeroa}).
The log divergence near $z=1$ is due to the $L_0=0$ states in $\left[  \left( V_{{\rm R}, 1}^{-{1\over 2},-{1\over 2}} \right)^{\otimes 2}\right]$, and is projected out in the modified amplitude ${\cal A}'$ (\ref{modamp}). The latter then evaluates to
\ie\label{modlambd}
& {\cal A}'\left[ \Psi' \otimes \left( V_{{\rm R}, 1}^{-{1\over 2},-{1\over 2}} \right)^{\otimes 2} \otimes \widehat\Psi_{\rm NS}^{(0)} \right] = 4{\A'^2 k_-^2}  \lambda' \, \langle k', F' | k,F\rangle
\fe
for some finite constant $\lambda'$ that depends on the choice of integration slice ${\cal S}_4\subset \widetilde{P}_{0,4}$.

Substituting $Y_{\rm R}^{(1)}=0$ and (\ref{pwnsm}), (\ref{modlambd}) into (\ref{simpfinaleqn}), we deduce the following equation for $Y_{\rm NS}^{(2)}$,
\ie\label{ynszzeq}
L_0 Y_{\rm NS}^{(2)} = 
- {\A'\over 4} k_-^2 \left( 2n \A'
+ Z^2  - 8\A' \ln r + 8 \A'\lambda' \right) \widehat\Psi_{\rm NS}^{(0)} +\cdots,
\fe
where $\cdots$ represents terms of other ghost structures or matter superconformal descendants. Recall that we are free to shift $Z^2$ by a constant via a spacetime coordinate transformation, which we can use to remove the $n$-independent constant $-8\A' \ln r + 8\A'\lambda'$ in (\ref{ynszzeq}). From (\ref{ynszzeq}) and moreover the first equation of (\ref{qbynseqn}), we deduce that $Y_{\rm NS}^{(2)}$ is of the form
\ie
Y_{\rm NS}^{(2)} = c\widetilde c e^{-\phi - \widetilde\phi}  e^{i(k_+X^++k_-X^-)} F^{(2)}(Z_i) {\cal O}_{\rm osc} + \cdots,
\fe
where the function $F^{(2)}$ obeys
\ie\label{fzzsug}
\left( 2k_+k_-+ \partial_{Z_i}\partial_{Z_i} - {4n\over \A'} \right) F^{(2)} = k_-^2 \left(2n\A' + Z^2 \right) F.
\fe
To summarize, we started with the $\mu^0$ maximally spinning NSNS string state of momentum $(k_+, k_-)$ in $(X^+,X^-)$ directions, and a wave function $F(Z_3,\cdots, Z_8)$ that is independent of $Z_1, Z_2$, and found that the NSNS string field receives a $\mu^2$ order correction to the $Z$-wave function of the form $\mu^2 F^{(2)}(Z_1,\cdots, Z_8)$ subject to (\ref{fzzsug}). 

To understand the meaning of (\ref{fzzsug}), let us compare it with the result obtained from quantizing the Green-Schwarz effective action \cite{Metsaev:2001bj, Metsaev2002a, Berenstein:2002jq}. 
In the pp-wave background (\ref{ppmetric}), the Green-Schwarz action reduces in a lightcone gauge to that of eight free massive bosons and eight free massive fermions, whose masses are equal and related to the RR flux. The spinning string states of our consideration correspond in the Green-Schwarz lightcone description to a closed string state with $n$ left-moving and $n$ right-moving oscillators excited in the first Fourier mode, as well as 4 fermionic zero mode excitations (out of the total 8 fermion zero modes). If we further excite $\ell$ zero modes of the $Z_i$'s, the dispersion relation of the string state is
\ie\label{gsdisp}
2k_+k_- = 2 |\mu k_-| (\ell+4) + {4n\over \A'} \sqrt{1 + {\A'^2 \mu^2 k_-^2}}.
\fe
Let $f(Z)$ be the wave function in the $Z_i$ zero modes of the string. We can re-express $\ell$ in terms of a harmonic oscillator Hamiltonian in the $Z_i$ directions acting on $f(Z)$, by substituting $2|\mu k_-|(\ell+4)\to - \partial_{Z_i}\partial_{Z_i} + \mu^2 k_-^2 Z^2$ in (\ref{gsdisp}). Expanding (\ref{gsdisp}) to order $\mu^2$ then gives
\ie\label{gskk}
\left( 2k_+k_- + \partial_{Z_i}\partial_{Z_i} - \mu^2 k_-^2 Z^2 - {4n\over \A'} -2 n \A' \mu^2 k_-^2 + {\cal O}(\mu^4) \right) f(Z)=0.
\fe
If we take $f(Z) = F(Z_3,\cdots Z_8) + \mu^2 F^{(2)}(Z_1,\cdots, Z_8) + {\cal O}(\mu^4)$, then (\ref{gskk}) agrees precisely with (\ref{fzzsug}). Thus, using the string field equation starting from (\ref{NSNSregge}), we have recovered a special case of the order $\mu^2$ dispersion relation of the maximally spinning string.

\section{Type IIB strings in $AdS_3\times S^3\times M_4$ with NSNS and RR flux}\label{sec:AdS3}

Now we turn to the primary example of interest, namely type IIB strings in $AdS_3\times S^3\times M_4$, where $M_4$ is either the $K3$ manifold or the four-torus $T^4$. These backgrounds arise as the near-horizon limit of the D1-D5-F1-NS5 system and play a central role in $AdS_3/{\rm CFT}_2$ dualities  \cite{Strominger1996,Maldacena1999e,Maldacena1998,David2002}. Furthermore, they provide a nontrivial but tractable laboratory for the application of our formalism for perturbative strings in mixed-flux backgrounds. This is because we can start with the $AdS_3\times S^3\times M_4$ background supported purely by NSNS fluxes, where the worldsheet theory in the NSR formalism admits a solvable CFT description that involves the ${\cal N}=1$ supersymmetric $SL(2,\mathbb{R})\times SU(2)$ WZW model \cite{Giveon1998,Kutasov1999,Maldacena2001a,Maldacena2002a}. We can then consider the deformed background by turning on RR 3-form fluxes. Note that while in the full non-perturbative string theory, both the NSNS and RR fluxes are quantized, at the level of string perturbation theory only the quantization of the NSNS flux is visible while the RR flux can be treated as a continuous parameter.

The type IIB supergravity solution for $AdS_3\times S^3\times M_4$ takes the form
\ie\label{adsssol}
& ds^2=R^2ds^2_{AdS_3}+R^2ds^2_{S^3}+ds^2_{M_4},
\\
& H_3=2qR^2(\omega_{AdS_3}+\omega_{S^3}),
\\
& F_3^{\rm RR}=2\sqrt{1-q^2}R^2(\omega_{AdS_3}+\omega_{S^3}),
\fe
where $\omega_{AdS_3}$ and $\omega_{S^3}$ are volume forms on $AdS_3$ and $S^3$ of unit radius, and $q$ is a parameter between 0 and 1. The quantization of NSNS flux is such that $R^2 = \A' k$ for a positive integer $k$. We begin with the case $q=1$, and turn on the RR flux by deforming to
\ie\label{qpara}
q=1-{\mu^2\over2},~~~~ R^2 = {\A' k\over 1-{\mu^2\over 2}},
\fe
so that the NSNS flux is held fixed. As already mentioned, since we will be working at the level of string perturbation theory, the quantization of RR flux is irrelevant, and we can treat $q$ as a continuous parameter.

\subsection{Worldsheet theory and the string spectrum in the purely NSNS background}

The type IIB string in the purely NSNS $AdS_3\times S^3\times M_4$ background is described in the NSR formalism by a worldsheet ``matter" SCFT that is the tensor product of the ${\cal N}=1$ $SL(2,\mathbb{R})_k\times SU(2)_k$ WZW model and a $c=6$ superconformal nonlinear sigma model on $M_4$, along with the superconformal ghosts and a suitable GSO projection. The supersymmetric $SL(2,\mathbb{R})_k$ WZW model, whose currents we denote $\mathfrak{J}^a$, $a=0,\pm$, can be constructed as the bosonic $SL(2,\mathbb{R})_{k+2}$ WZW model whose currents are denoted $J^a$, together with three free fermions $\psi^a$ and their anti-holomorphic counterparts. The supersymmetric $SU(2)_k$ WZW model is constructed from the bosonic $SU(2)_{k-2}$ WZW model whose currents are denoted $K^{a'}$, $a'=0,\pm$, together with three free fermions $\chi^{a'}$ and their anti-holomorphic counterparts. The detailed conventions are summarized in Appendix \ref{adsconv}.

As building blocks of vertex operators, the current algebra primaries of the bosonic $SL(2,\mathbb{R})_{k+2}$ are denoted $V^{sl}_{j,m,\bar m}$, whose conformal weights are $h=\tilde h = -{j(j-1)\over k}$, whereas the current algebra primaries of the bosonic $SU(2)_{k-2}$ are denoted $V^{su}_{j',m',\bar m'}$ with weight $h=\tilde h = {j'(j'+1)\over k}$. Physical string states will be described by (delta-function) normalizable vertex operators that involve unitary representations of the $SL(2,\mathbb{R})$ as well as states related by spectral flow \cite{Maldacena2001a}. It is important here that spectral flow is that of the ${\cal N}=1$ $SL(2,\mathbb{R})_k$. In this paper, we will restrict to principal discrete representations $\mathcal{D}_j^{\pm}$ of $SL(2,\mathbb{R})$ with no spectral flow. The RR deformation of the background will be described using non-normalizable vertex operators that involve finite dimensional non-unitary representations of $SL(2,\mathbb{R})$, such as the fundamental ($j=-{1\over 2}$) or the adjoint ($j=-1$) representation.

For simplicity we will analyze massive pulsating string modes, described by $(-1,-1)$ picture NSNS vertex operators of the form $c\widetilde c  e^{-\phi-\tilde\phi} {\cal V}_{j,j',n}$, with
\ie\label{pulsatingvert}
{\cal V}_{j,j',n} = {1\over  2n!(k-2j+2)_n} \psi^-\widetilde\psi^- (J_{-1}^-)^n (\widetilde J_{-1}^-)^n V^{sl}_{j,j,j} V^{su}_{j',j',j'}V_{M_4},
\fe
where $V_{M_4}$ is a bosonic superconformal primary vertex operator of the $M_4$ worldsheet CFT. For example, in the case of $M_4=T^4$, we can take $V_{M_4}=e^{ip\cdot X}$ where $X_i$'s for $i=1,2,3,4$ correspond to four compact $U(1)$ bosons of $T^4$ and the $p_i$ are the corresponding momenta. The matter part of ${\cal V}_{j,j',n}$ is a superconformal primary which is the lowest weight state in the spin $j-n-1$ representation with respect to both the left and the right $SL(2,\mathbb{R})$ global symmetry, with ${\mathfrak{J}}^3_0 = \widetilde {\mathfrak{J}}^3_0 = j-n-1$. Denoting the weights of $V_{M_4}$ as $(h_{\rm int},h_{\rm int})$, the BRST-closure/mass-shell condition amounts to
\ie\label{massjj}
-{j(j-1)\over k} + n + {j'(j'+1)\over k} +h_{\rm int} = 0,
\fe
or equivalently
\ie
(j+j')(j-j'-1) = (n+h_{\rm int})k.
\fe
Note that the spacetime energy is $\DD=2(j-n-1)$ whereas the angular momentum quantum number on the $S^3$ is $j'$. The positivity of spacetime energy requires $j-n-1>0$. This also leads, by spectral flow symmetry, to an upper bound on $j$ so that
\ie\label{bound}
n+1<j<{k\over2}-n+1,
\fe 
which is stronger than the no-ghost bound \cite{Evans:1998qu}. We also have $0\leq j'\leq {k-2\over2}$ for half-integer $j'$ of $SU(2)_{k-2}$. To facilitate a comparison with the BMN limit \cite{Berenstein:2002jq}, we take $h_{\rm int}=0$ and define
\ie\label{eq:AdS3LightconeMomenta}
p^+ = {j-n-1+j' \over k},~~~p^- = j-n-1- j',
\fe
where $H_{l.c.}=2p^-$ can be identified with the light-cone Hamiltonian. The mass-shell condition (\ref{massjj}) can be written as
\ie\label{masspp}
(kp^++n+1)\left(p^- +n\right) = nk.
\fe
The BMN limit corresponds to taking $j,j',k\gg 1$, holding $j-j'$ and $n$ fixed, so that $p^+$ and $p^-$ are both of order 1. In this limit (\ref{masspp}) reduces to
\ie
p^+ p^- = n ( 1- p^+),
\fe
which agrees with the string spectrum in the purely NSNS pp-wave background, in the case where only oscillators of the first Fourier mode on the string are excited \cite{Berenstein:2002jq,Dei:2018yth}.

\subsection{Turning on RR fluxes}

We now turn on RR flux by taking $\mu$ to be nonzero in (\ref{adsssol}), (\ref{qpara}), while fixing the NSNS flux number $k$. Consequently the overall radius $R$ of $AdS_3\times S^3$ is deformed according to (\ref{qpara}). The deformed supergravity solution, a priori valid in the large $k$ limit, corresponds to a solution to the string field equation of the form (\ref{psizero}), (\ref{psivv}), with
\ie\label{vvdefs}
& {\cal O}_{{\rm R},1} = {1\over 4\pi}\sqrt{2\over k} \left[ S_+^{\A\A'} \Theta_+ \widetilde S_+^{\db\db'}  \widetilde\Theta_+ \right] (V^{sl}_{j=-{1\over 2}})_{\A\dot\B} (V^{su}_{j'={1\over 2}})_{\A'\dot\B'} .
\fe
Here we denote by $\Theta_\pm$ and $\widetilde\Theta_\pm$ the holomorphic and anti-holomorphic spin fields in the superconformal nonlinear sigma model on $M_4$, where the subscript $\pm$ labels their chirality. Note that only the combination $\Theta_+\widetilde\Theta_+$ appears in the first order RR string field.\footnote{This is also consistent with the RR deformation operator for the corresponding pp-wave background viewed as a deformation of Minkowskian spacetime.} $S$ and $\widetilde S$ are holomorphic and anti-holomorphic spin fields constructed from the bosonization of the 6 free fermions as part of the ${\cal N}=1$ $SL(2,\mathbb{R})\times SU(2)$ WZW model. The components of $S$ are denoted $S_\pm^{\A\A'}$ where $\A$ and $\A'$ are doublet indices of the $sl(2,\mathbb{R})\simeq so(1,2)$ and $su(2)\simeq so(3)$ respectively, and the subscript $\pm$ indicates the chirality of the spin field with respect to the $SO(1,5)$ acting on the 6 fermions. 

The order $\mu^2$ NSNS string field can be solved from (\ref{vwsep}) and (\ref{wnseqn}). In this case, the RHS of (\ref{wnseqn}) vanishes, and we can simply set $W^{(2)}_{\rm NS}$ to zero. The matter CFT operator ${\cal O}_{{\rm NS},2}$ appearing in (\ref{vnssketch}) takes the form
\ie\label{matternsco}
& {\cal O}_{{\rm NS},2} ={1\over 4\pi}\left[ A^{sl} \psi^a \widetilde\psi^{b} (V^{sl}_{j=-1})_{ab} + A^{su} \chi^{a'} \widetilde\chi^{b'} (V^{su}_{j'=1})_{a'b'} \right] + \cdots,
\fe
where $\cdots$ represents higher weight operators. The coefficients $A^{sl}$ and $A^{su}$ can be computed from (\ref{simplaftem}), with the result
\ie\label{aarrex}
A^{sl} = r^{-{4\over k}} C^{sl}_{-{1\over 2},-{1\over 2},-1}, ~~~A^{su} = r^{{4\over k}} C^{su}_{{1\over 2},{1\over 2},1},
\fe
where $r$ is the disc radius parameter in the definition of the string field bracket as in (\ref{simplaftem}). Here we denote by $C^{sl}_{j_1, j_2, j_3}$ and $C^{su}_{j_1' ,j_2' ,j_3'}$ the structure constants of primaries of the corresponding spins in the bosonic $SL(2,\mathbb{R})_{k+2}$ and $SU(2)_{k-2}$ WZW models, respectively. In the case of interest, the fundamental-fundamental-adjoint structure constants are given by \cite{Zamolodchikov:1986bd}
\ie\label{OPEcoeffs}
{}C^{su}_{{1\over 2},{1\over 2},1}=&{\Gamma(1-2/k)\over\Gamma(2/k)}\sqrt{\Gamma(1/k)\Gamma(3/k)\over\Gamma(1-1/k)\Gamma(1-3/k)},
~~~~ C^{sl}_{-{1\over 2},-{1\over 2},-1}=&{4\over3}\left(C^{su}_{{1\over 2},{1\over 2},1}\right)^{-1}.
\fe
Both $C^{sl}_{-{1\over 2},-{1\over 2},-1}$ and $C^{su}_{{1\over 2},{1\over 2},1}$ approach ${2\over\sqrt{3}}$ in the large radius limit $k\to\infty$. In this limit, the picture-raised version of the NSNS string field $e^{-\phi -\widetilde\phi} {\cal O}_{{\rm NS},2}$ is proportional to the kinetic term in the WZW Lagrangian density, as expected from the radius deformation of $AdS_3\times S^3$. From the point of view of worldsheet CFT deformation (\ref{rrdef}), the parameter $r$ in (\ref{matternsco}), (\ref{aarrex}) again plays the role of a short distance cutoff, as the non-marginality of ${\cal O}_{{\rm NS},2}$ serves to cancel the Weyl anomaly due to a pair of colliding RR deformation operators.

For later convenience, we also write down the picture-raised RR string field at order $\mu$,
\ie{}
V_{{\rm R},1}^{{1\over 2},{1\over 2}} =& c\widetilde c e^{{\phi\over 2} + {\widetilde\phi\over 2}} {\cal O}_{{\rm R},1}^\uparrow\\
 =& {1\over 8\pi k} \sqrt{2\over k} \, c\widetilde c e^{{\phi\over 2} + {\widetilde\phi\over 2}} \left[ S_-^{\A\A'} \Theta_+ \widetilde S_-^{\db\db'}  \widetilde\Theta_+ \right]
\\
&\times\Big[  (J_{-1})_\A{}^\C (\widetilde J_{-1})_\db{}^\dd (V^{sl}_{j=-{1\over 2}})_{\C\dd} (V^{su}_{j'={1\over 2}})_{\A'\dot\B'} +  (J_{-1})_\A{}^\C (\widetilde K_{-1})_{\db'}{}^{\dd'} (V^{sl}_{j=-{1\over 2}})_{\C\db} (V^{su}_{j'={1\over 2}})_{\A'\dd'} 
\\
& +(K_{-1})_{\A'}{}^{\C'} (\widetilde J_{-1})_\db{}^\dd (V^{sl}_{j=-{1\over 2}})_{\A\dd} (V^{su}_{j'={1\over 2}})_{\C'\dot\B'} +  (K_{-1})_{\A'}{}^{\C'} (\widetilde K_{-1})_{\db'}{}^{\dd'} (V^{sl}_{j=-{1\over 2}})_{\A\db} (V^{su}_{j'={1\over 2}})_{\C'\dd'}  \Big]
\\
&
+ ({\rm terms~involving~}\Theta_-{\rm ~or~}\widetilde\Theta_-).
\fe

The terms involving $\Theta_-$ or $\widetilde\Theta_-$ will not be important for the computations below, as they will not contribute to the mass correction to the string states of our consideration at order $\mu^2$.

\subsection{Corrections to the pulsating string states}

We now study the linearized string field $\widehat\Psi$ describing the pulsating string state in the RR deformed background, whose order $\mu^0$ term is $\widehat\Psi_{\rm NS}^{(0)} = c\widetilde c e^{-\phi - \widetilde\phi} {\cal V}_{j,j',n}$. The mass correction is captured by (\ref{simpfinaleqn}), with $W^{(2)}_{\rm NS}=0$ as already mentioned. Choosing $\Psi'$ in (\ref{simpfinaleqn}) to be $\overline{\widehat\Psi_{\rm NS}^{(0)}}$, the 4-point amplitude ${\cal A}$ with a pair of RR background field insertions is computed from the integrated four point function
\ie\label{eq:IntegratedFourPoint}
 {1\over 2} {\cal A} \left[ \overline{\widehat\Psi_{\rm NS}^{(0)}} \otimes \left( V_{{\rm R}, 1}^{-{1\over 2}, -{1\over 2}} \right)^{\otimes 2} \otimes \widehat\Psi_{\rm NS}^{(0)} \right] 
=&-{\pi}\int d^2 z \,{|1-z|^{1\over 2} |z|} \left\langle\overline{\cal V}_{j,j',n}(\infty) {\cal O}_{{\rm R},1}(1) {\cal O}^\uparrow_{{\rm R},1}(z,\bar z) {\cal V}_{j,j',n}(0)\right\rangle\\ 
\equiv& - {\pi\over 2} \int d^2z f_{j,j',n}(z,\bar z),
\fe
where we have moved the two PCOs to $z, \bar z$ to convert the string field insertion $V_{{\rm R},1}^{-{1\over 2},-{1\over 2}}(z,\bar z)$ to the $({1\over 2},{1\over 2})$ picture. The moduli $z$-integral has a power divergence near $z=1$, having to do with the relevant part of ${\cal O}_{{\rm NS},2}$ appearing in the OPE of the two RR deformation string fields. Such a power divergence is standard in string perturbation theory. In SFT, it can be cured by separating the relevant operator from the string field bracket involving $\left( V_{{\rm R}, 1}^{-{1\over 2}, -{1\over 2}} \right)^{\otimes 2}$ in solving the string field equation. The end result is equivalent to simply throwing away the power divergence.

Note that there is no log divergence from the region near $z=0$ or $\infty$, indicating that ${\cal A}$ is the same as the modified amplitude ${\cal A}'$, and that we can set $Y^{(1)}_{\rm R}=0$ in (\ref{simpfinaleqn}). Together with $W^{(2)}_{\rm NS}=0$ and (\ref{eq:IntegratedFourPoint}), (\ref{simpfinaleqn}) reduces to
\ie{}
\mu^2\left\langle \widehat\Psi_{\rm NS}^{(0)} \right| \left. c_0\widetilde c_0 L_0 Y_{\rm NS}^{(2)}\right\rangle = - \delta h \left\langle \widehat\Psi_{\rm NS}^{(0)} \,\right|\left.\widehat\Psi_{\rm NS}^{(0)} \right\rangle,
\fe
where $\delta h$ or the corresponding mass squared correction $\delta m^2$ are given by
\ie\label{eq:AdS3RRDeformation}
\delta h = {\A'\over 4} \delta m^2  = - {\pi\over 2} \mu^2 \int d^2z f_{j,j',n}(z,\bar z).
\fe
From the viewpoint of deforming the worldsheet CFT by (\ref{rrdef}), we may view $\delta h$ as an anomalous weight of the vertex operator ${\cal V}_{j,j',n}$ at order $\mu^2$, hence the notation.

The explicit evaluation of $f_{j,j',n}$ involves two steps. First, due to the appearance of $SL(2,\mathbb{R})_{k+2}$ current algebra descendants in $V_{{\rm R},1}^{{1\over2},{1\over2}}$ as well as in ${\cal V}_{j,j',n}$, we need to apply Ward identities to reduce the computation to that of correlators of current algebra primaries, as described in Appendix \ref{WardId}. Next, correlators of current algebra primaries are computed using the Knizhnik-Zamolodchikov (KZ) equation \cite{Knizhnik:1984nr}, which is performed in Appendices \ref{SU4pt}, \ref{SL4pt}. Using these results, the four point function is written as 
\ie\label{fourpointfunction}
 f_{j,j',n}(z,\bar z)=&{\mu^2\over 4\pi^2k^2}{\E^{\A'\mu'}\E^{\dot\B'\dot\nu'}}\left(\E^{\A\mu} z -\D^{\A,-}\D^{\mu,+} \right)\left(\E^{\dot\B\dot\nu} \bar z -\D^{\dot\B,-}\D^{\dot\nu,+} \right)
\\
&\times\bigg\{t_{a\mu}^{~~\gamma}t_{b\dot\nu}^{~~\dot\delta}\left[\mathbb{X}^a(z)\right]_{\gamma\alpha}^{\lambda\tau}\left[\mathbb{X}^b(\bar z)\right]_{\dot\delta\dot\beta}^{\dot\lambda\dot\tau}G^{sl,j}_{\lambda\dot\lambda;\tau\dot\tau}(z,\bar z)G^{su,j'}_{\mu'\dot\nu';\alpha'\dot\beta'}(z,\bar z)\\
&+t_{a\mu}^{~~\gamma}T_{a'\dot\nu'}^{~~~\dot\delta'}\left[\mathbb{X}^a(z)\right]_{\gamma\alpha}^{\lambda\tau}\left[\mathbb{W}(\bar z)\right]_{\dot\nu\dot\beta}^{\dot\lambda\dot\tau}[\mathbb{Y}^{a'}(\bar z)]_{\dot\delta'\dot\beta'}^{\dot\rho'\dot\sigma'}G^{sl,j}_{\lambda\dot\lambda;\tau\dot\tau}(z,\bar z)G^{su,j'}_{\mu'\dot\rho';\alpha'\dot\sigma'}(z,\bar z)\\
&+T_{a'\mu'}^{~~~\gamma'}t_{a\dot\nu}^{~~\dot\delta}[\mathbb{W}(z)]_{\mu\alpha}^{\lambda\tau}[\mathbb{X}^a(\bar z)]_{\dot\delta\dot\beta}^{\dot\lambda\dot\tau}[\mathbb{Y}^{a'}(z)]_{\gamma'\alpha'}^{\rho'\sigma'}G^{sl,j}_{\lambda\dot\lambda;\tau\dot\tau}(z,\bar z)G^{su,j'}_{\rho'\dot\nu';\sigma;\dot\beta'}(z,\bar z)\\
&+T_{a'\mu'}^{~~~\gamma'}T_{b'\dot\nu'}^{~~~\dot\delta'}[\mathbb{W}(z)]_{\mu\alpha}^{\lambda\tau}[\mathbb{W}(\bar z)]_{\dot\nu\dot\beta}^{\dot\lambda\dot\tau}[\mathbb{Y}^{a'}(z)]_{\gamma'\alpha'}^{\rho'\sigma'}[\mathbb{Y}^{b'}(\bar z)]_{\dot\delta'\dot\beta'}^{\dot\rho'\dot\sigma'}G^{sl,j}_{\lambda\dot\lambda;\tau\dot\tau}(z,\bar z)G^{su,j'}_{\rho'\dot\rho';\sigma'\dot\sigma'}(z,\bar z)\bigg\}.
\fe
Here, $G^{sl,j}$ and $G^{su,j'}$ are four point functions of $SL(2,\mathbb{R})_{k+2}$ and $SU(2)_{k-2}$ current algebra primaries given by (\ref{slfourpt}) and (\ref{sufourpt}), with $(z_1, z_2)$ set to $(z,1)$. $\mathbb{W},\mathbb{X},\mathbb{Y}$ are matrices acting on the fundamental indices of the four point functions given by (\ref{A1def}), (\ref{A2def}), and (\ref{Bdef}), also with the substitution $(z_1, z_2) \to (z,1)$. Note that $V_{M_4}$ in ${\cal V}_{j,j',n}$ contains only bosonic excitations in $M_4$, whereas the terms in $V^{{1\over 2},{1\over 2}}_{{\rm R},1}$ that involve bosonic excitations in $M_4$ must come with spin fields $\Theta_-$ or $\widetilde\Theta_-$ and would not contribute to the correlator in question. Thus, the $M_4$ part of the correlator in (\ref{fourpointfunction}) is trivial.

\subsubsection{BPS states}

An important consistency check is the case in which the pulsating string vertex operators ${\cal V}_{j,j',n}$ are chiral primaries, namely $n=0,~ j=j'+1, ~h_{\rm int}=0$ \cite{Dabholkar2009}, describing BPS supergraviton states. Such states should receive no mass corrections under the RR deformation. Instead of demonstrating the preservation of spacetime supersymmetry under the deformation by solving for the gauge symmetries perserving the background, we can verify the absence of mass corrections to the BPS states at order $\mu^2$, by evaluating the integral on the RHS of (\ref{eq:AdS3RRDeformation}) including a counter term that subtracts off the power divergence at $z=1$,  
\ie\label{bpscheck}
\delta h&=- {\pi\over 2} \int_{\bC} d^2 z \left[ f_{j'+1,j',0}(z,\bar z) - {B_{j'}\over |1-z|^{2+{4\over k}}}\right] \\
&=-{\mu^2(j'+1)^2\over 8\pi k^2}\int_{\bC} d^2 z\bigg\{ {1\over |z(1-z)|^2}\bigg[|z|^2 g^{sl,j'+1}_{++;--}(z,\bar z)g^{su,j'}_{--;++}(z,\bar z)\\
&~~~ - z g^{sl,j'+1}_{+-;-+}(z,\bar z)g^{su,j'}_{-+;+-}(z,\bar z)-\bar z g^{sl,j'+1}_{-+;+-}(z,\bar z)g^{su,j'}_{+-;-+}(z,\bar z)\\
&~~~ +g^{sl,j'+1}_{--;++}(z,\bar z)g^{su,j'}_{++;--}(z,\bar z)\bigg]
-{4^{1+{4\over k}}\cos({\pi\over k})^2\csc({2\pi j'\over k})\csc({2\pi(j'+1)\over k})\Gamma({1\over 2}+{1\over k})^4\over \Gamma(1-{2j'\over k})^2\Gamma({2(j'+1)\over k})^2|1-z|^{2+{4\over k}}}\bigg\} \\
&\dot{=} ~0,
\fe 
where the last equality is confirmed numerically. Here, the counter term coefficient $B_{j'}$ is determined by inspecting the $z\to 1$ limit of the 4-point function. $g^{sl,j}$ and $g^{su,j'}$ are four point functions of $SL(2,\mathbb{R})_{k+2}$ and $SU(2)_{k-2}$ current algebra primaries given by (\ref{sl4pt}) and (\ref{su4pt}).

There are two cases of (\ref{bpscheck}), where $j=j'+1$ is either 1 or ${k\over 2}$, that require special treatment. Writing $g^{sl,j}$ as a linear combination of $SL(2)$ current algebra conformal blocks (\ref{sl4pt}), the coefficient $h_-^{24}$ diverges at $j=1$, whereas $h_+^{24}$ diverges at $j={k\over2}$. However, $j=1$ is excluded by the bound (\ref{bound}). On the other hand, $j={k\over2}$ is allowed, and while the 4-point function in the integrand of (\ref{eq:IntegratedFourPoint}) diverges in this case, the divergent coefficient is multiplied by an integral in the cross ratio that vanishes, thus a more careful regularization is required. While $j$ a priori takes discrete values, as constrained by the mass-shell condition, we can consider $AdS_3\times S^3\times T^4$ where the $T^4$ is large and a near-BPS supergraviton mode that carries small momentum on the $T^4$, corresponding to a string vertex operator of the form (\ref{pulsatingvert}) with $j'={k\over 2}-1$, $n=0$, and small $h_{\rm int}>0$. We can then take the limit $h_{\rm int}\to 0$, in which $j$ approaches ${k\over2}$ from above. One can indeed verify numerically that $\delta h$ converges to zero in this limit.

\subsubsection{Non-BPS states}
\label{nonbpsstaessec}

\begin{table}[h!]
\centering
\begin{tabular}{l||c|c|c|c|c}
\backslashbox{$k$}{$j'$} & $0$ & $1\over 2$ & $1$ & ${3\over 2}$ & $2$ \\
\hline\hline
$7$ & $4.51353$ & $7.7253$ &  & &\\
\hline
$8$ & $2.61214$ & $3.18173$ &  $5.03926$ & $38.0435$ &\\
\hline
$9$ & $1.97318$ & $2.21068$ & $2.76008$ & $4.25035$ & $15.9923$\\
\end{tabular}
\caption{The leading anomalous weight in units of $\mu^2$ for some small values of the level $k$ (or equivalently, the $AdS$ radius) at $n=1$ (with $h_{\rm int} = 0$). For each value of $k$, the range of $j'$ considered eventually truncates once the $SL(2,\mathbb{R})_{k+2}$ spin $j$ fixed by the mass-shell condition (\ref{massjj}) saturates or exceeds the upper bound $k\over 2$ (cf (\ref{bound})).}
\label{tab:SmallKAnomalousWeight}
\end{table}

Now we turn to the mass correction due to RR flux at order $\mu^2$ for the non-BPS pulsating string states, described by the vertex operators ${\cal V}_{j,j',n}$, where $n\geq 1$, $0\leq j'\leq {k\over 2} - 1$, with $j$ determined by the mass-shell condition (\ref{massjj}) and subject to the bound (\ref{bound}). Here we compute $\delta h$ by numerically integrating the 4-point function in (\ref{eq:IntegratedFourPoint}). Some explicit numerical values of $\delta h/\mu^2$ for small values of $k$ are shown in Table \ref{tab:SmallKAnomalousWeight}.

\begin{figure}[h!]
\centering
\subfloat{
\includegraphics[width=.45\textwidth]{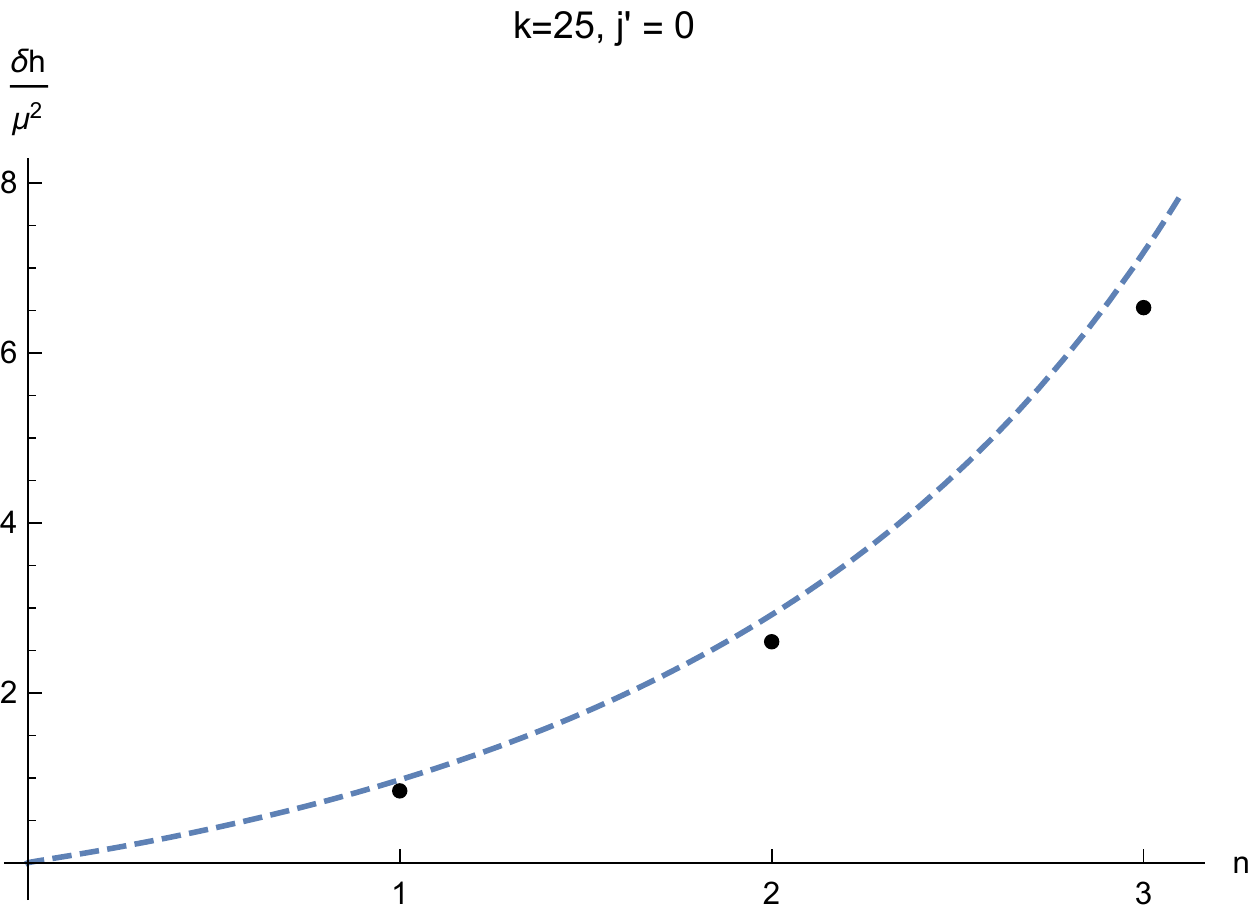}
}
\subfloat{
\includegraphics[width=.45\textwidth]{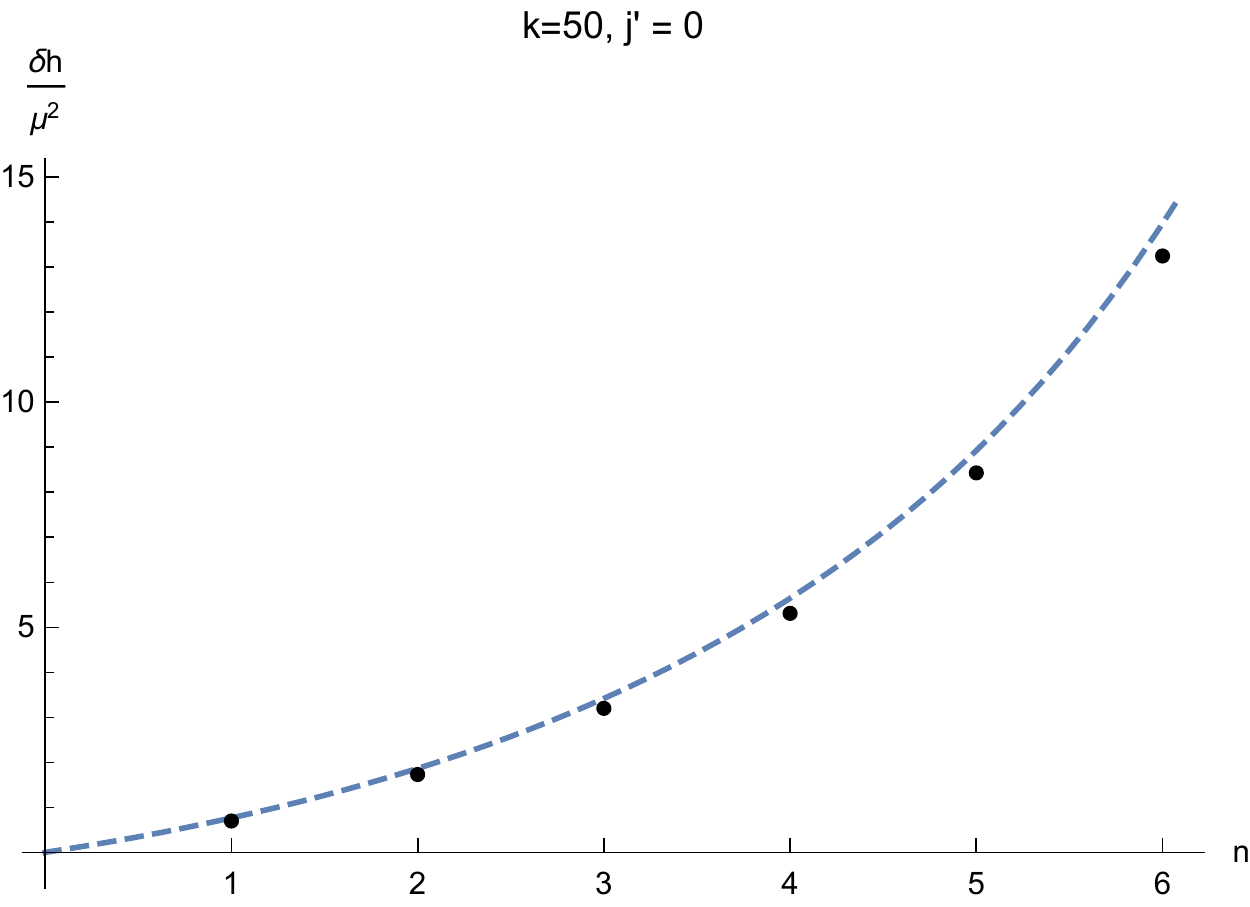}
}
\\
\subfloat{
\includegraphics[width=.45\textwidth]{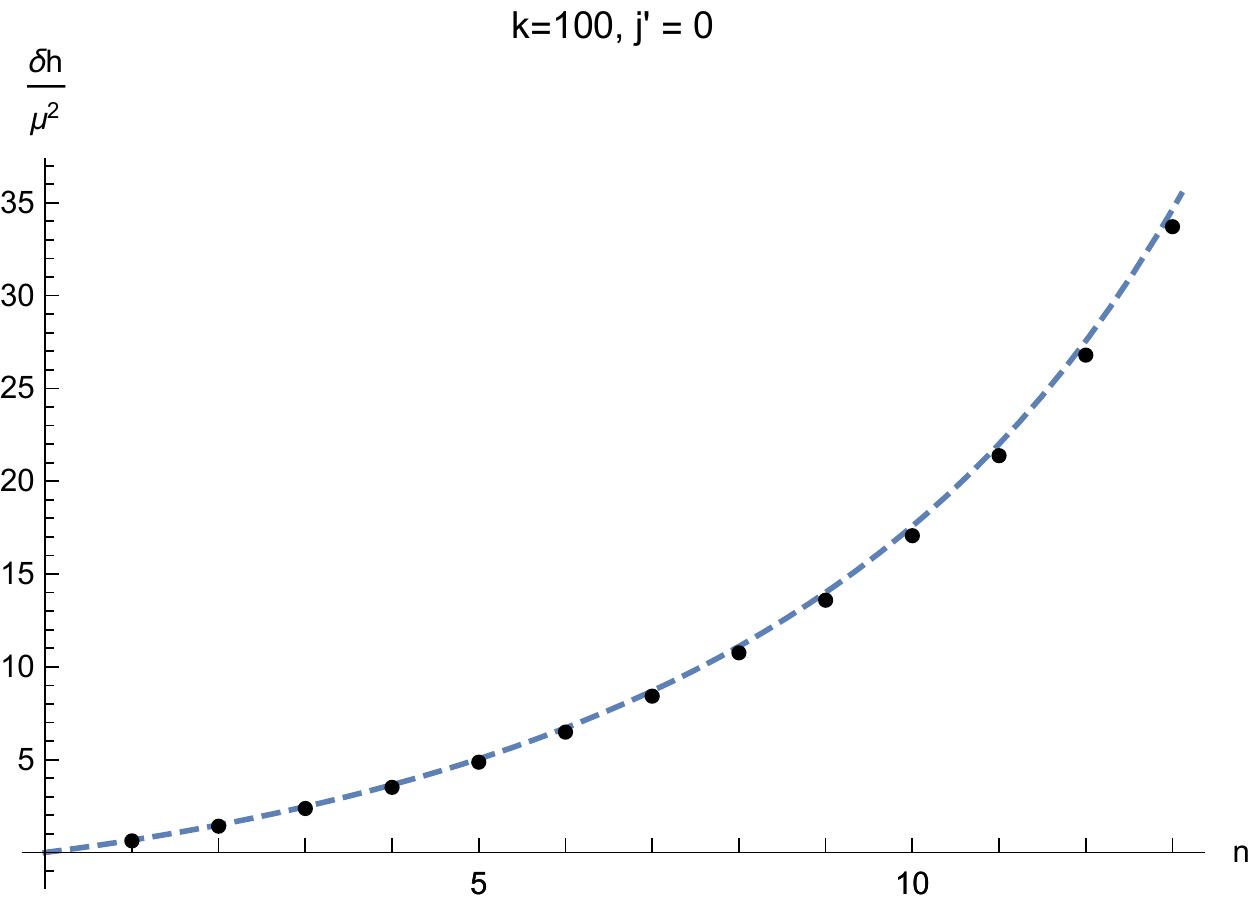}
}
\subfloat{
\includegraphics[width=.55\textwidth]{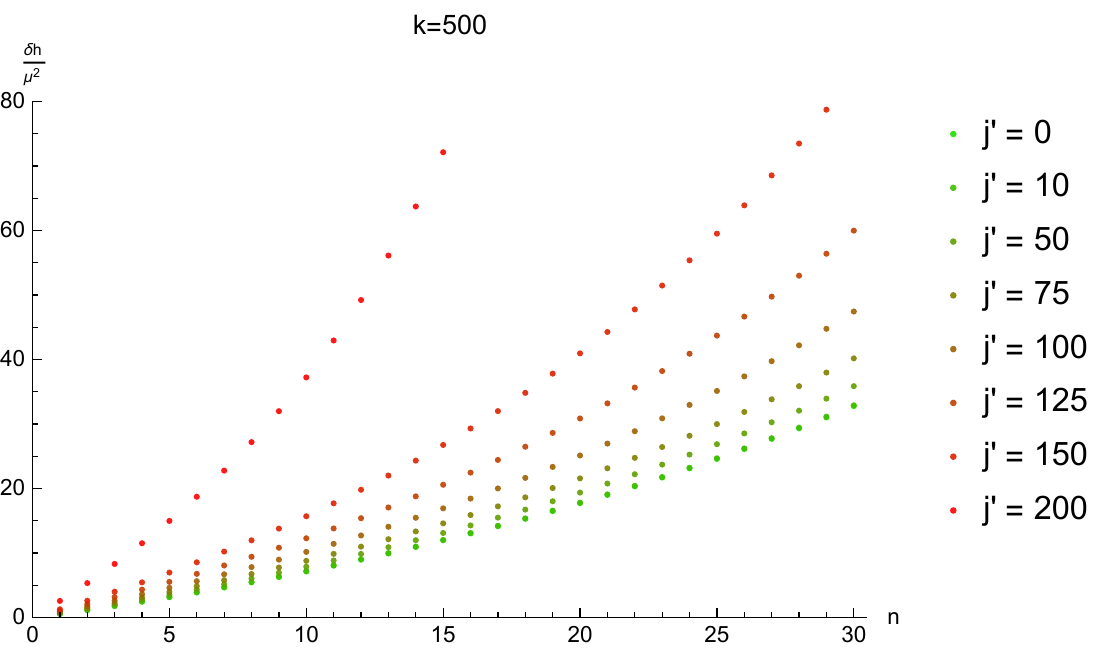}
}
\caption{The leading anomalous weight $\delta h$ in units of $\mu^2$ as a function of the oscillator number $n$ (with $h_{\rm int}=0$). In the first three plots, the blue dashed curve represents the anomalous weight inferred from the semiclassical quantization of the pulsating string solution (\ref{semiclassic}), while the black dots denote the quantum result. In the bottom-right plot, we have plotted the quantum anomalous weights (colored dots) for various $S^3$ angular momentum quantum numbers $j'$.}\label{fig:FiniteDeformation}
\end{figure}

In Figure \ref{fig:FiniteDeformation}, we plot our numerical results of $\delta h/\mu^2$ for larger fixed level $k$, and for fixed $SU(2)$ spin $j'$, as a function of the oscillator number $n$. The results are compared to those obtained from semiclassical quantization, as we analyze in the next subsection.

The spectrum of $j$ is of course discrete once $h_{\rm int}$ is specified. However, since at this order $h_{\rm int}$ enters the computation in a trivial way, the $SL(2,\mathbb{R})_{k+2}$ spin $j$ can essentially be regarded as a continuous variable. As $j$ approaches $k\over 2$ from below, we observe a divergence in the anomalous weight, which has the interpretation of a short pulsating string becoming long \cite{Maldacena2001a}. Indeed, we will see this feature explicitly in the semiclssical regime (\ref{semiclassic}) in the next subsection.

\subsection{The semiclassical limit}

The classical pulsating string solutions in $AdS_3$ with mixed flux were studied in \cite{Hernandez:2018gcd}. At the classical level, the RR flux only affects the relation between the radius $R$ and the $H$-flux number $k$, namely ${k\alpha'\over R^2}=q=1-{\mu^2\over2}$. In the conformal gauge, the relevant worldsheet Lagrangian is
\ie
{\cal L}={R^2\over 4\pi\alpha'}\bigg[(1+r^2)(-\dot t^2+t'^2)+{1\over1+r^2}(\dot r^2-r'^2)+r^2(\dot\phi^2-\phi'^2)+{2k\alpha'r^2\over R^2}(\dot t\phi'-\dot\phi t')\bigg],
\fe
where $(t,r,\phi)$ are the worldsheet fields corresponding to $AdS_3$ global coordinates ($r\equiv\sinh \rho$), $\cdot$ stands for derivative with respect to the worldsheet time $\tau$, and $'$ the derivative with respect to worldsheet spatial coordinate $\sigma$. We will set $\alpha'=1$ in the rest of this section.  A (non-spinning) pulsating string is represented by
\ie\label{trphiw}
t=t(\T),~r=r(\T),~\phi=\s.
\fe
It is easy to see that the $\D \phi$ component of the equation of motion (EOM) and the Virasoro constraint $T_{\sigma\tau}=0$ are trivially satisfied by (\ref{trphiw}). Nontrivial components of the EOM read\footnote{Note that if there are internal excitations, the RHS of the last line of (\ref{deltrtt}) would be replaced by $-{4h\over R^2}$, where $h$ is  the ``internal weight" (in the language of the quantum theory).}
\ie\label{deltrtt}
{}&\D t : \partial_\T\left(-2(1+r^2)\dot t+{2kr^2\over R^2}\right)=0,
\\
&\D r : 2r\dot t^2-{2r\dot r^2\over(1+r^2)^2}+{2\ddot r\over1+r^2}+2r-{4rk\over R^2}\dot t=0,
\\
&T_{\s\s}=T_{\T\T} \propto -(1+r^2)\dot t^2+{\dot r^2\over1+r^2}+r^2=0.
\fe
The $\D t$-EOM gives $\dot t={1\over1+r^2}\left({kr^2\over R^2}+{\DD\over R^2} \right)$, where $\DD$ is the spacetime energy
\ie
\DD={R^2\over4\pi}\int_0^{2\pi}d\s{\D L\over\D\dot t}={R^2\over4\pi}\int_0^{2\pi}d\s \left[2(1+r^2)\dot t-{2kr^2\phi'\over R^2}\right].
\fe
The Virasoro constraint can then be written as
\ie\label{conservationEq}
\dot r^2=C_1r^4+C_2r^2+C_3,
\fe
with $C_1=\left({k\over R^2}\right)^2-1$, $C_2={2k\DD\over R^4}-1$, $C_3={\DD^2\over R^4}$.

The canonical momentum conjugate to $r$ is given by $p_r= R^2 {\dot r\over1+r^2}$. We can now proceed with semiclassical quantization, by the Bohr-Sommerfeld quantization condition 
\ie\label{prrphas}
\oint p_r dr= 2R^2\int_0^{r_0}dr{\sqrt{C_1 r^4+C_2r^2+C_3}\over1+r^2}=2\pi n,~~~n\in\mathbb{Z}.
\fe
The integral on the LHS is over the orbit in the $(r, p_r)$ phase space. Note that as the pulsating string goes from shrinking to expanding at $r=0$, the sign of $p_r$ flips. $r_0$ is the maximal value of $r$, where $\dot r=0$. 

\begin{figure}[h!]
\centering
\subfloat{
\includegraphics[width=.95\textwidth]{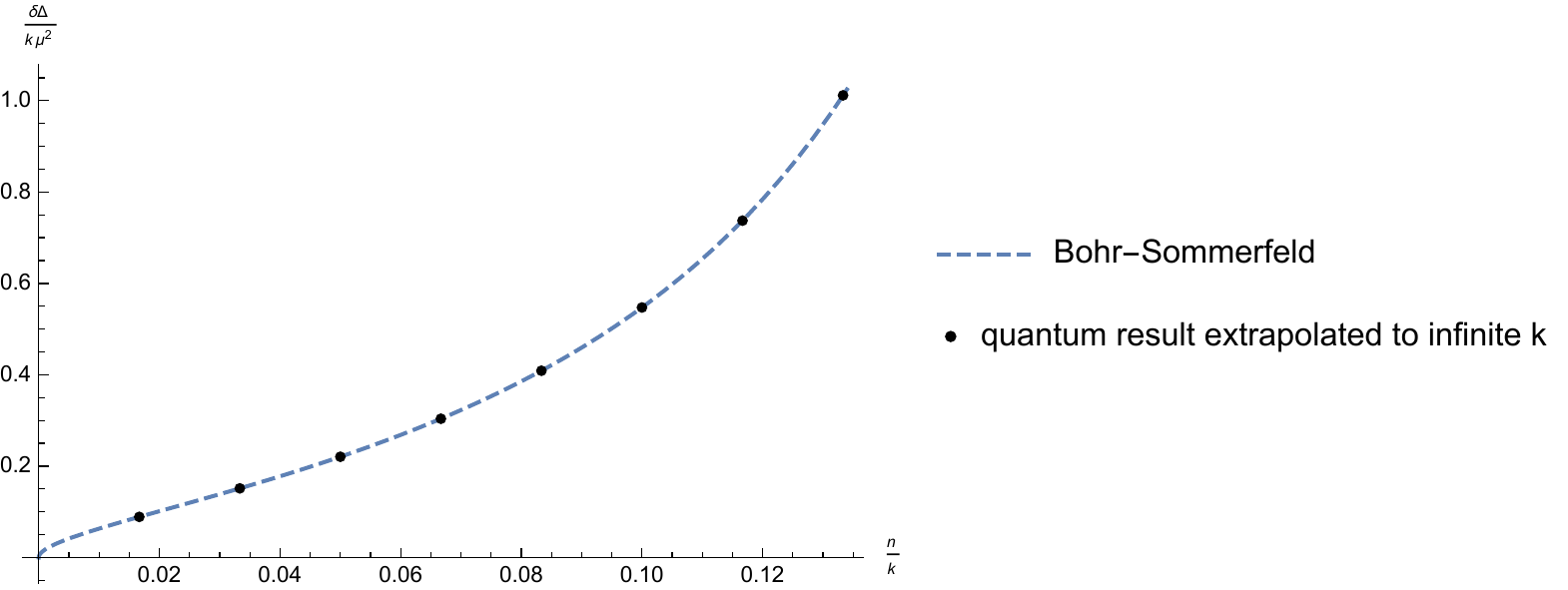}
}
\caption{RR correction to the spacetime energy $\D\DD$ in units of $k\mu^2$, numerically extrapolated to infinite level $k$ at fixed ${n\over k}$ (black dots), compared to the semiclassical result (\ref{semiclassic}) (dashed blue curve).}\label{fig:semiclassics}
\end{figure}

The $r$-integral in (\ref{prrphas}) can be evaluated by changing the variable $r\equiv r_0\,\text{sn}(u\tau,v)$, where $u=\sqrt{-C_2+\sqrt{C_2^2-4C_1C_3}\over2}$, $v={-C_2-\sqrt{C_2^2-4C_1C_3}\over -C_2+\sqrt{C_2^2-4C_1C_3}}$, and $\text{sn}(u,v)$ is the Jacobi elliptic sn function. The resulting quantization condition is
\ie\label{BSquant}
n={uR^2\over2r_0^2\pi}\left(-r_0^2 E(\text{am}(2K(v),v),v)+(1+r_0^2)(-2vK(v)+(r_0^2+v)\Pi(-r_0^2,\text{am}(2K(v),v),v) \right),
\fe
where $K$, $E$, $\Pi$, am are complete elliptic integral of the first kind, elliptic integral of the second kind, incomplete elliptic integral, and Jacobi amplitude respectively. Expanding this relation in $\mu$ to order $\mu^2$ leads to the following energy spectrum
\ie\label{semiclassic}
{}&\DD=\DD_0+\D\DD,
\\
&\DD_0=-2n+2\sqrt{nk},~~\D\DD=\mu^2\left[{\sqrt{nk}\over2}+{2nk-3n\sqrt{nk}\over2(2\sqrt{n}-\sqrt{k})^2}\right].
\fe
Note the divergence of $\delta\Delta$ at $n = {k\over 4}$, which corresponds to $j\sim {k\over 2}$ in the semiclassical regime (cf (\ref{massjj})), where the short strings become long. This can be compared to the order $\mu^2$ quantum results computed in the previous section. In the semiclassical limit, the anomalous weight $\delta h$ is related to the correction to spacetime energy $\delta \Delta$ in the following way (in the case $j'=0$ with no internal excitation)
\ie
\D h={j(j-1)\over k}-n\approx{1\over k}\left({(\DD_0+\D\DD)\over2}+n\right)^2-n\approx\D\DD\sqrt{n\over k}.
\fe
With this identification, a numerical extrapolation of our quantum results to the $k\to \infty$ limit is in accurate agreement with the semiclassical quantization, as shown in Figure \ref{fig:semiclassics}.\footnote{For each value of $n\over k$, the extrapolation was performed by computing the correction to the spacetime energy for values of $k$ ranging from $300$ to $2100$ in steps of $300$ and fitting $\delta\Delta\over k\mu^2$ to a polynomial of degree 5 in $k^{-1}$.}

\subsection{The BMN limit}

We can compare our results with the previously known spectrum of strings in pp-wave background, obtained from the Penrose/BMN limit of $AdS_3\times S^3$ \cite{Berenstein:2002jq}. Writing the $AdS_3\times S^3$ metric as
\ie\label{eq:AdS3S3Metric}
R^2(ds^2_{AdS_3}+ds^2_{S^3}) =& R^2(-\cosh^2\rho~ dt^2 + d\rho^2+\sinh^2\rho ~d\phi_1^2 +\cos^2\theta~ d\psi^2 + d\theta^2+\sin^2\theta ~d\phi_2^2),
\fe
the Penrose limit is defined by zooming in near a null geodesic corresponding to a massless particle moving in $\psi$ direction on the $S^3$ while staying at the center of $AdS_3$. To this end we introduce coordinates $\tilde{x}^{\pm} = {1\over \sqrt{2}}(t\pm \psi)$, define the rescaled coordinates
\ie\label{eq:BMNLimit}
x^+ ={1\over\sqrt{2}} \tilde{x}^+,~x^- = \sqrt{2} R^2\tilde{x}^-,~\rho ={r\over R},~\theta = {y\over R},
\fe
and take the limit $R\to \infty$. The resulting background is that of the pp-wave geometry
\ie
R^2(ds^2_{AdS_3}+ds^2_{S^3})\to -2 dx^+dx^- -(r^2+y^2)(dx^+)^2+dr^2+r^2d\phi_1^2+dy^2+y^2d\phi_2^2,
\fe
supported by three-form fluxes
\ie{}
& H_3 \to 2 q\left(dx^+\wedge dx^1\wedge dx^2+dx^+\wedge dx^3\wedge dx^4\right),
\\
& F_3^{\rm RR}\to 2\sqrt{1-q^2}\left(dx^+\wedge dx^1\wedge dx^2 + dx^+\wedge dx^3\wedge dx^4\right),
\fe
where we have redefined $re^{i\phi_1} = x_1+ix_2$ and $ye^{i\phi_2} = x_3+ix_4$. The spectrum of strings in the pp-wave background is naturally described in terms of the lightcone momenta
\ie\label{pdeltaj}
p^+ = {\Delta + J_\psi \over R^2}, ~~~ p^- = \Delta - J_\psi,
\fe
where $\Delta$ and $J_\psi$ are the charges associated with the isometries generated by $\partial_t$ and $\partial_\psi$. (\ref{pdeltaj}) is the same as (\ref{eq:AdS3LightconeMomenta}) in the purely NSNS case. In the mixed flux case, we must keep in mind that the radius $R$ is given by (\ref{qpara}).

The full spectrum of type IIB strings in the above pp-wave background has been determined from the lightcone quantization of the Green-Schwarz effective string action \cite{Berenstein:2002jq,Dei:2018yth}, and recently reproduced in the hybrid formalism \cite{Berkovits1999, Eberhardt2018}. Here we briefly recap the results following the notation of \cite{Dei:2018yth}. In the lightcone gauge, the Green-Schwarz action of strings in the pp-wave background is that of free massive bosons and fermions on the worldsheet. Focusing on bosonic oscillator excitations transverse to the $M_4$ directions, the lightcone Hamiltonian can be written, up to a ground state energy, as the following
\ie
H_{l.c.}=\sum_{i=1,2}\sum_{n\in\mathbb{Z}}\left(\omega_n^+a^{i,+\dag}_na^{i,+}_n+\omega_n^-a^{i,-\dag}_na^{i,-}_n \right),
\fe
where $a_n^{i,\pm}$ and $(a_n^{i,\pm})^\dagger$ are the annihilation and creation operators for a pair of complex bosons, and satisfy the commutation relations
\ie{}
[a^{i,\pm}_n,(a^{i,\pm}_m)^\dag]=\delta_{nm},~~~~[a^{i,\pm}_n,(a^{i,\mp}_m)^\dag]=0.
\fe
The oscillator energies $\omega_n^\pm$ are given by the dispersion relation
\ie
\omega_n^{\pm}=\sqrt{1\pm{2qn\over p^+}+{n^2\over (p^+)^2}}.
\fe
In addition, the total momentum along the string is constrained to be zero (level matching condition).

It was shown in \cite{Dei:2018yth} that at the purely NSNS point, the $SL(2,\mathbb{R})$ oscillators $J^\pm_{-n}$, $\tilde{J}^\pm_{-n}$ in the NSR formalism of the worldsheet theory correspond in the BMN limit to $(a^{1,\pm}_n)^\dag$, $(a^{1,\mp}_{-n})^\dag$ among the above described oscillators of the lightcone Green-Schwarz string. Therefore, the pulsating string states (\ref{pulsatingvert}) correspond to, in the BMN limit, the Green-Schwarz string with $n$ $(a^{1,-}_1)^\dag$ and $n$ $(a^{1,+}_{-1})^\dag$ excitations. This obeys the level matching condition. Note that, importantly, $(a^{1,-}_1)^\dag$ and $(a^{1,+}_{-1})^\dag$ create oscillator modes of the {\it same} lightcone energy, $\omega^-_1=\omega^+_{-1}$.

After turning on RR flux, the spectrum of the pulsating strings in the pp-wave is determined by
\ie\label{ppspect}
2p^-=H_{l.c.}=2n\omega^-_1=2n\sqrt{1-{2q\over p^+}+{1\over (p^+)^2}}=2n\left[{1-p^+\over p^+}+{\mu^2\over2(1-p^+)}\right] +\mathcal{O}(\mu^4).
\fe

\begin{figure}[h!]
\centering
\subfloat{
\includegraphics[width=.49\textwidth]{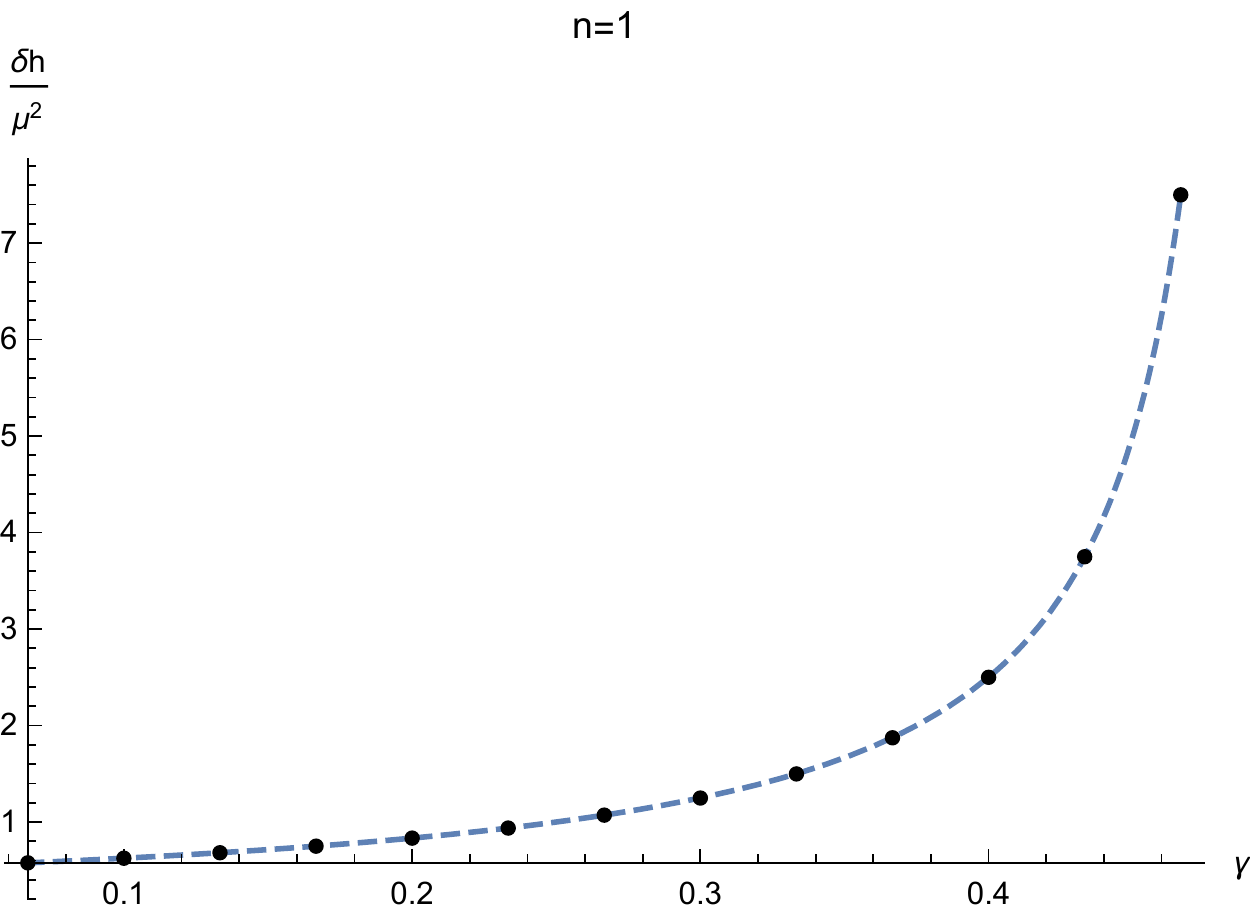}
}
\subfloat{
\includegraphics[width=.49\textwidth]{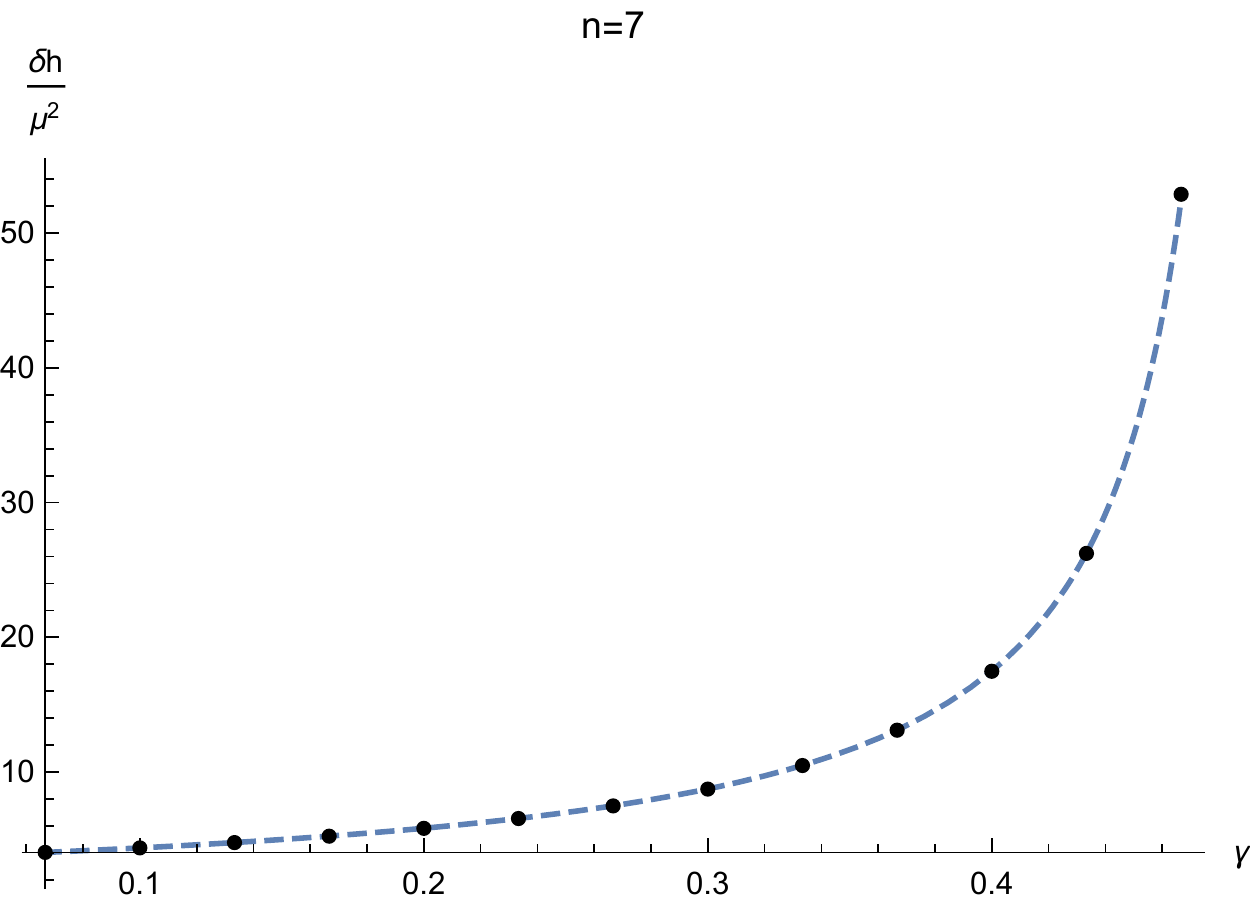}
}
\caption{The order $\mu^2$ anomalous weight $\delta h$ extrapolated to infinite level $k$ at fixed $\gamma={j'\over k}$ (black dots), compared to that of the known pp-wave string spectrum (\ref{eq:PPExpectation}) (dashed blue curve), at two sample values of the oscillator level $n$.}\label{fig:ppDeformation}
\end{figure}

In the previous section, we have computed numerically the order $\mu^2$ correction to the spectrum of the pulsating strings in $AdS_3\times S^3$ of finite radius, through the anomalous weight $\delta h$. The lightcone momentum (\ref{pdeltaj}) is expressed in terms of the quantum numbers $j, j', n$ as
\ie
p^+= \left(1-{\mu^2\over2}\right) {j-n-1+j'\over k} ,~~~~ p^- = j-n-1-j'.
\fe
The anomalous weight is related to the lightcone momenta $p^+,p^-$ as follows
\ie\label{eq:Ads3ppAnomalousWeight}
\left({k p^+ \over 1-{\mu^2\over 2}} +n+1\right)(p^-+n) = (n + \delta h) k.
\fe
The BMN limit amounts to sending $k\to \infty$ while keeping $\gamma\equiv j'/k$ finite and $n$ fixed. The mass-shell condition then sets $j=j'+1+{n+\delta h\over 2\gamma} + {\cal O}(k^{-1})$. We perform the computation described in section \ref{nonbpsstaessec} for $\delta h$ at order $\mu^2$ as a function of $k$ and $\gamma$, for fixed $n$, and take the BMN limit by numerically extrapolating to $k=\infty$. The result as a function of $\gamma$ fits accurately to
\ie\label{eq:PPExpectation}
\delta h=n\mu^2\left({\gamma\over 1-2\gamma}+{1\over2} \right)+\mathcal{O}(\mu^4),
\fe
as demonstrated in Figure \ref{fig:ppDeformation}.\footnote{For each value of $\gamma$, the extrapolation was performed by computing the anomalous weight for values of $k$ ranging from $300$ to $1500$ in steps of $150$ and fitting $\delta h\over \mu^2$ to a polynomial of degree 6 in $k^{-1}$.} This is precisely in agreement with the Green-Schwarz string spectrum (\ref{ppspect}).

\section{Discussion}
\label{discsec}

In a rough sense, Ramond-Ramond flux backgrounds can be treated perturbatively as a nonlocal but nonetheless conformally invariant deformation of the worldsheet CFT, within the Neveu-Schwarz-Ramond formalism \cite{Berenstein:1999jq,Berenstein:1999ip}. We redefined such a prescription using the framework of closed superstring field theory. 
As a basic check of our formalism, we considered type IIB strings in pp-wave background supported by 5-form RR flux, viewed as a deformation of Minkowskian spacetime, and analyzed corrections to the dispersion relation of a family of spinning string states to second order in the RR flux. This is a rather delicate example as the deformation changes the asymptotics and thereby normalizability of the string wave functions. While our result agrees with a special case of the previously known string spectrum in pp-wave obtained from the Green-Schwarz formalism, various features of the string field solution illustrated through our derivation may provide useful guidance for analyzing, say, the spectrum of type IIB string theory in $AdS_5\times S^5$ viewed as a deformation of Minkowskian spacetime.

We then considered a more nontrivial example, namely type IIB strings in $AdS_3\times S^3\times M_4$ supported by mixed 3-form fluxes, viewed as an RR deformation of the purely NSNS background. The general structure of the perturbative string field solution in this case is in fact much simpler than the pp-wave case, as the RR deformation does not change the spacetime geometry qualitatively, although the detailed computation requires some $SL(2,\mathbb{R})\times SU(2)$ current algebra gymnastics. Our main result is the second order RR correction to the mass spectrum of the pulsating strings at finite $AdS$ radius. To the best of our knowledge, this result has not been attainable in alternative formulations of superstring perturbation theory in RR backgrounds. 
Our result passes two consistency checks. In the limit of large $AdS$ radius and large oscillation quantum number $n$, the mass corrections agree with the Bohr-Sommerfeld quantization of a pulsating string. In the BMN limit, our result is in agreement with the previously known spectrum of strings in the pp-wave background with mixed fluxes.

Let us briefly comment on other attempts of formulating superstring perturbation theory in RR backgrounds. The Green-Schwarz formalism \cite{Green1984a, Grisaru:1985fv, Wulff:2016tju} is based on a worldsheet effective action constrained by spacetime local super-Poincar\'e symmetry. Such an effective string action is subject to the ambiguity of deformation by higher dimensional operators \cite{Aharony:2013ipa, Hellerman:2014cba}, and the rule of quantization is a priori unclear except in simple backgrounds, such as in Minkowskian or pp-wave spacetimes in which the string action happens to be free in a suitable gauge. Nonetheless, substantial progress has been made in quantizing the GS action for strings in $AdS_5\times S^5$ \cite{Frolov:2002av, Callan:2003xr, Callan:2004uv, Roiban:2007jf, Roiban:2007dq, Arutyunov:2009ga, Giombi:2009gd}. Under the assumption that an appropriate quantization of the GS action for strings in $AdS_3\times S^3$ will respect the integrability observed of the classical string in this background, the spectrum of quantum strings (in particular in the case of mixed 3-form flux) has been analyzed in \cite{Hoare:2013ida, Hoare:2013lja, Hoare:2013pma, Babichenko:2014yaa, Borsato:2015mma, Borsato:2016xns, Baggio:2017kza, OhlssonSax:2018hgc}. It would be very interesting to compare our results to those obtained from integrability.

The hybrid \cite{Berkovits1999} and pure spinor \cite{Berkovits2000b} formalisms are based on the BRST framework as in the NSR formalism, while replacing the worldsheet CFT by ones defined through local actions that appear to accommodate RR flux backgrounds and respect spacetime symmetries. The main difficulty with these formalisms, in our opinion, is that the worldsheet ghost-matter coupling is not fully defined at the quantum level, which appears to be the main obstacle in its application to, for instance, strings in $AdS$ at finite radius.\footnote{This is perhaps not an essential obstacle. Intriguing results suggestive of the physical string spectrum have been found in the matter supergroup sigma model recently \cite{Eberhardt:2018vho}.} Note that in the type IIB pp-wave background supported by mixed 3-form fluxes, the ghost-matter coupling is absent, and indeed the string spectrum  in this case has been recovered from the hybrid formalism in \cite{Eberhardt2018}.

The investigation of the string spectrum in $AdS_3\times S^3\times M_4$ with mixed flux is particularly interesting due to the role of the latter in the $AdS_3/CFT_2$ correspondence \cite{Maldacena1999e, Aharony:1999ti, Larsen:1999uk}. So far, for simplicity, we have restricted our attention to a special class of short string states, namely the pulsating strings (that may also be orbiting on the $S^3$). Another class of short string states of interest are those on the leading Regge trajectory. In the presence of NSNS $H$-flux, as was pointed out in \cite{Loewy:2002gf}, the folded spinning string solutions of \cite{Gubser:2002tv} are unstable, and the leading Regge trajectory string states have more intricate classical limits. At the purely NSNS point, such states have been studied in \cite{Ferreira:2017pgt}. It should be a straightforward extension of the computation performed in the present paper to analyze the leading RR corrections to all short string states starting from the purely NSNS background.

The purely NSNS background has the distinguishing feature that it admits long strings \cite{Seiberg:1999xz, Maldacena2001a} that can be viewed as scattering states in global $AdS_3$, whose energy spectrum forms a continuum above a gap. When RR fluxes are turned on, the long string spectrum is expected to become discrete. The simple perturbative treatment of the RR flux in this paper is not directly applicable in computing the mass spectrum of the now-discrete long string states.

In this paper we merely made use of the classical version of SFT, for the sake of a consistent treatment of off-shell deformation operators in the worldsheet theory. It is conceivable that, at the level of string spectrum and tree level string perturbation theory, the SFT framework is ultimately not necessary and one may be able to formulate the RR background deformation in terms of a nonlocal conformal worldsheet theory.
The nonlocality under consideration here is in a mild sense; that is, they should be ``local up to BRST exact terms" \cite{Berenstein:1999ip, Berenstein:1999jq}.
Other types of nonlocal deformations on the worldsheet have been considered previously in  \cite{Aharony2001,Aharony2002a,Aharony2002}. Nonetheless, from the point of view of effective string theory \cite{Polchinski:1991ax, Dubovsky:2012sh, Aharony:2013ipa, Hellerman:2014cba, Dubovsky:2015zey}, there is no a priori reason to expect the worldsheet theory of perturbative strings to be described by a local quantum field theory. It is perhaps no coincidence that holographic confining backgrounds for QCD-like strings generally involve RR fluxes.

\section*{Acknowledgments}

We would like to thank Ofer Aharony, Bruno Balthazar, Lorenz Eberhardt, Rajesh Gopakumar, Simeon Hellerman, Daniel Jafferis, Roji Pius, Victor Rodriguez, Eva Silverstein, and Bogdan Stefanski for enlightening conversations and correspondences. We are especially grateful to Ashoke Sen for extensive comments and discussions on an early draft. We thank the participants of the 2018 meeting of the Simons collaboration on the non-perturbative bootstrap for many interesting conversations and the Caltech high energy theory group for hospitality, where some of this work was performed. SC and XY thank KITP for its hospitality during the finishing stage of this work. This work is supported in part by a Simons Investigator Award from the Simons Foundation, by the Simons Collaboration Grant on the Non-Perturbative Bootstrap, and by DOE grant $\text{DE-SC00007870}$. MC is supported in part by Samsung Scholarship. SC is supported in part by the Natural Sciences and Engineering Research Council of Canada via a PGS D fellowship. 

\appendix
\section{Worldsheet CFT conventions for $AdS_3\times S^3$}\label{adsconv}
The worldsheet CFT for the purely NSNS $AdS_3\times S^3$ background can be constructed from the bosonic $SL(2,\mathbb{R})_{k+2}$ and $SU(2)_{k-2}$ WZW models and $SL(2,\mathbb{R})_{-2}$ and $SU(2)_{2}$ free fermions. The relevant OPEs are 
\ie{}
&J^a(z)J^b(w)\sim {{k+2\over 2}q_{sl}^{ab}\over(z-w)^2}+{f_c^{ab}J^c(w)\over z-w},
\\
&\psi^a(z)\psi^b(w)\sim {q_{sl}^{ab}\over z-w},
\fe
where $J^a$ is the bosonic $SL(2,\mathbb{R})$ current. $q_{sl}^{ab}$ is the Killing form given by $-q^{sl}_{00}=2q^{sl}_{+-}=1$, and $f_c^{ab}$ are $SL(2)$ structure constants $f_+^{0+}=-f_-^{0-}=1, f_0^{+-}=-2$. In the $SU(2)$ case, where $K^{a'}$ is the bosonic current,
\ie{}
&K^{a'}(z)K^{b'}(w)\sim {{k-2\over 2}q_{su}^{a'b'}\over(z-w)^2}+{g_c^{a'b'}K^{c'}(w)\over z-w},
\\
&\chi^{a'}(z)\chi^{b'}(w)\sim {q_{su}^{a'b'}\over z-w},
\fe
with $q^{su}_{00}=2q^{su}_{+-}=1$, $g_+^{0+}=-g_{-}^{0-}=1, g_0^{+-}=2$.

The spin fields $S_{\pm}^{\A\A'}$ transform in the fundamentals under both $SL(2,\mathbb{R})_{-2}$ and $SU(2)_{2}$. Fundamental indices are contracted with epsilon tensors $\E_{+-}=\E^{+-}=1$ as $r^\A=\E^{\A\B}r_\B,r_\A=r^\B\E_{\B\A}$. The relevant OPEs are
\ie{}
S_+^{\A\A'}(z)S_-^{\B\B'}(w)\sim&{\E^{\A\B}\E^{\A'\B'}\over(z-w)^{3/4}},
\\
S_{\pm}^{\A\A'}(z)S_{\pm}^{\B\B'}(w)\sim&{\sqrt{2}(\E^{\A'\B'}t_a^{\A\B}\psi^a-\E^{\A\B}T_{a'}^{\A'\B'}\chi^{a'})\over(z-w)^{1/4}},
\\
\chi^{a'}(z)S_{\pm}^{\A\A'}(w)\sim&{\sqrt{2}T^{a'\A'}_{~~~\B'}S_{\mp}^{\A\B'}\over(z-w)^{1/2}},
\\
\psi^{a}(z)S_{\pm}^{\A\A'}(w)\sim&{\sqrt{2}t^{a\A}_{~~~\B}S_{\mp}^{\B\A'}\over(z-w)^{1/2}}.
\fe
Here $t^{a~~\B}_{~\A}=\left(t^{a}\right)^{~\B}_{\A}$ and $T^{a'~\B'}_{~\A'}=\left(T^{a'}\right)_{\A'}^{~\B'}$ are $SL(2,\mathbb{R})$ and $SU(2)$ matrices in fundamental representations. In the standard basis, they are given by $t^0_{~\pm\mp}={1\over2}, t^{\pm}_{~\pm\pm}=\mp i$, and $T^0_{~\pm\mp}={1\over2}, T^{\pm}_{~\pm\pm}=\mp1$. The fermionic OPEs lead to the following four point function
\ie\label{adsfermi4pt}
\langle\psi^a(z_1)\psi^b(z_2)S_+^{\A\A'}(z_3)S_-^{\B\B'}(z_4)\rangle={q_{sl}^{ab}\E^{\A\B}\E^{\A'\B'}(z_{13}z_{24})^{1/2}\over z_{12}(z_{14}z_{23})^{1/2}z_{34}^{3/4}}-{2\E^{\A'\B'}t^{a\A}_{~~\C}t^{b\B\C}z_{34}^{1/4}\over(z_{13}z_{14}z_{23}z_{24})^{1/2}},
\fe
where we have also used the relation
\ie
t^{a\B}_{~~\C}t^{b\A\C}=-{q_{sl}^{ab}\E^{\A\B}\over2}+t^{a\A}_{~~\C}t^{b\B\C}.
\fe
We furthermore define
\ie
J^{\A\B}=\sqrt{2}J^at_a^{~\A\B},~~K^{\A'\B'}=\sqrt{2}K^{a'}T_{a'}^{~\A'\B'}.
\fe

\section{Current algebra Ward identities}\label{WardId}

In this section we derive some useful Ward identities that reduce correlation functions of $SL(2,\mathbb{R})$ current algebra descendants to those of the current algebra primaries.
We start by stating the commutation relations among $SL(2,\bR)_{k+2}$ generators,
\ie{}\label{eq:SL2Algebra}
[J^3_n,J^3_m] =& -{k+2\over 2}n\delta_{m+n,0}\\
[J^3_n,J_m^{\pm}] =& \pm J_{m+n}^{\pm}\\
[J_n^+,J_m^-] =& -2J^{3}_{m+n}+(k+2)n\delta_{m+n,0}.
\fe
A current algebra primary $V^{sl}$ obeys commutation relation of the form
\ie{}
[J^a_n,V^{sl}(z_i)] =& z^n (t^a_i)^T V^{sl}(z_i),
\fe
where $t^a_i$ are the $SL(2,\mathbb{R})$ generator matrices; the subscript $i$ is meant to denote that they are acting on the current algebra primary at $z_i$.
Focusing on the holomorphic part, the basic correlators that will be needed for the computation of the 4-point function $f_{j,j',n}$ in (\ref{eq:IntegratedFourPoint}) are of the form
\ie
(W)_{\alpha\beta}\equiv& \langle(J^+_{-1})^nV^{sl}_{j,-j}(\infty)V^{sl}_{-1/2,\alpha}(z_1)V^{sl}_{-1/2,\beta}(z_2)(J^-_{-1})^nV^{sl}_{j,j}(0)\rangle\\
(X^a)_{\alpha\beta}\equiv &\langle(J^+_{-1})^nV^{sl}_{j,-j}(\infty)J^a_{-1}V^{sl}_{-1/2,\alpha}(z_1)V^{sl}_{-1/2,\beta}(z_2)(J^-_{-1})^nV^{sl}_{j,j}(0)\rangle.
\fe
Here the operator $V_{j,j}^{sl}$ corresponds to the lowest weight state $\ket{j,j}$ of $\mathcal{D}_j^+$, which in particular obeys $J_0^-\ket{j,j} = 0$.

We begin by evaluating the following chain of current algebra generators that will be useful in the computations that follow:
\ie{}
(J^+_{+1})^m(J^-_{-1})^n\ket{j,j} =&{n!\over (n-m)!}{(k-2j+2)_n\over (k-2j+2)_{n-m}}(J^-_{-1})^{n-m}\ket{j,j}\\
=& (n-m+1)_m(k-2j+n-m+2)_m(J^-_{-1})^{n-m}\ket{j,j}\\
\equiv & f_{n,m}(J^-_{-1})^{n-m}\ket{j,j}
\fe
where $(a)_n = {\Gamma(a+n)\over \Gamma(a)}$ is the Pochhammer symbol. In particular, $f_{n,n} = n! (k-2j+2)_n$ will enter the normalization of the pulsating string vertex operators.

Next, we compute the correlator $W$ (in intermediate steps of computations we will omit the fundamental indices on $V^{sl}_{-1/2}$, with the understanding that the matrices $t_i$ act on $V^{sl}_{-1/2}(z_i)$)
\ie\label{A1def}
W=&\bra{j,j}(J^+_{+1})^nV^{sl}_{-1/2}(z_1)V^{sl}_{-1/2}(z_2)(J^-_{-1})^n\ket{j,j}\\ 
=& \bra{j,j}V^{sl}_{-1/2}(z_1)V^{sl}_{-1/2}(z_2)(J^+_{+1})^n(J^-_{-1})^n\ket{j,j}\\
&+n(z_1 t_1^-+z_2 t_2^-)\bra{j,j}V^{sl}_{-1/2}(z_1)V^{sl}_{-1/2}(z_2)(J^+_{+1})^{n-1}(J^-_{-1})^n\ket{j,j}\\
&+n(n-1)z_1z_2t^-_1t^-_2\bra{j,j}V^{sl}_{-1/2}(z_1)V^{sl}_{-1/2}(z_2)(J^+_{+1})^{n-2}(J^-_{-1})^n\ket{j,j}\\
=& \left(f_{n,n}-nf_{n,n-1}(z_1t_1^-+z_2t_2^-)(z_1^{-1}t_1^++z_2^{-1}t_2^+)+2n(n-1)f_{n,n-2}t_1^{-}t_2^{-}t_1^{+}t_2^+\right)\\
&\times\bra{j,j}V^{sl}_{-1/2}(z_1)V^{sl}_{-1/2}(z_2)\ket{j,j}
\\
\equiv&\mathbb{W}(z_1,z_2)\bra{j,j}V^{sl}_{-1/2}(z_1)V^{sl}_{-1/2}(z_2)\ket{j,j},
\fe
where we have used $(J^\pm_0)^2 V^{sl}_{-1/2,m} = 0$, and $\mathbb{W}(z_1,z_2)$ is defined as a matrix acting on the holomorphic $SL(2,\mathbb{R})$ spinor indices. Explicitly writing out the fundamental indices, we have
\ie
(W)_{\alpha\beta} =& \left[\mathbb{W}(z_1,z_2)\right]^{\rho\sigma}_{\alpha\beta}\bra{j,j}V^{sl}_{-1/2,\rho}(z_1)V^{sl}_{-1/2,\sigma}(z_2)\ket{j,j}
\fe
The matrix acting on the anti-holomorphic spinor indices is analogously defined.

To compute $X^a$, we use the contour representation of $J_{-1}^a$ at $z_2$, and deform the contour around $z_1$, giving
\ie
X^a=&\langle (J^+_{-1})^nV^{sl}_{j,-j}(\infty)J^a_{-1}V^{sl}_{-1/2}(z_1)V^{sl}_{-1/2}(z_2)(J^-_{-1})^nV^{sl}_{j,j}(0)\rangle\\
=& \langle\oint_{C(\infty)}{dy\over 2\pi i}\sum_{\ell=0}^\infty z_1^{\ell}y^{-(\ell+1)}J^a(y)(J^+_{-1})^nV^{sl}_{j,-j}(\infty)V^{sl}_{-1/2}(z_1)V^{sl}_{-1/2}(z_2)(J^-_{-1})^nV^{sl}_{j,j}(0)\rangle\\
& - \langle\oint_{C(z_2)}{dy\over 2\pi i}(y-z_1)^{-1}(y-z_2)^{-1}(J^+_{-1})^nV^{sl}_{j,-j}(\infty)V^{sl}_{-1/2}(z_1)J^a_0V^{sl}_{-1/2}(z_2)(J^-_{-1})^nV^{sl}_{j,j}(0)\rangle\\
&+\langle\oint_{C(0)}{dy\over 2\pi i}\sum_{\ell=0}^\infty z_1^{-(\ell+1)}y^{\ell}(J^+_{-1})^nV^{sl}_{j,-j}(\infty)V^{sl}_{-1/2}(z_1)V^{sl}_{-1/2}(z_2)J^a(y)(J^-_{-1})^nV^{sl}_{j,j}(0)\rangle\\
=& \sum_{\ell=0}^{n-1}z_1^{\ell}\bra{j,j}(J^+_{+1})^nJ^a_{-(\ell+1)}V^{sl}_{-1/2}(z_1)V^{sl}_{-1/2}(z_2)(J^-_{-1})^n\ket{j,j}\\
&+z_{12}^{-1}(t^a_2)^T\bra{j,j}(J^+_{+1})^nV^{sl}_{-1/2}(z_1)V^{sl}_{-1/2}(z_2)(J^-_{-1})^n\ket{j,j}\\
&+\sum_{\ell=0}^nz_1^{-(\ell+1)}\bra{j,j}(J^+_{+1})^nV^{sl}_{-1/2}(z_1)V^{sl}_{-1/2}(z_2)J^a_\ell(J^-_{-1})^n\ket{j,j}.
\fe
Using the lowest weight property of $|j,j\rangle$ one finds that the first sum on the RHS reduces to the $\ell=0$ term, whereas the second sum reduces to $\ell=0,1$ terms, giving
\ie\label{eq:A2Correlator}
X^a=&\bra{j,j}(J^+_{+1})^nJ^a_{-1}V_{-1/2}(z_1)V_{-1/2}(z_2)(J^-_{-1})^n\ket{j,j}\\
& + z_{12}^{-1}(t_2^a)^T\bra{j,j}(J^+_{+1})^nV_{-1/2}(z_1)V_{-1/2}(z_2)(J^-_{-1})^n\ket{j,j}\\
&+z_1^{-1}\bra{j,j}(J^+_{+1})^nV_{-1/2}(z_1)V_{-1/2}(z_2)J^a_0(J^-_{-1})^n\ket{j,j}\\
&+z_1^{-2}\bra{j,j}(J^+_{+1})^nV_{-1/2}(z_1)V_{-1/2}(z_2)J^a_1(J^-_{-1})^n\ket{j,j}.
\fe
The first term, for instance, can be evaluated as
\ie
&\bra{j,j}(J^+_{+1})^nJ^a_{-1}V^{sl}_{-1/2}(z_1)V^{sl}_{-1/2}(z_2)(J^-_{-1})^n\ket{j,j}\\
=& \delta^{a,-}n(k-2j+n+1)\bra{j,j}(J^{+}_{+1})^{n-1}V^{sl}_{-1/2}(z_1)V^{sl}_{-1/2}(z_2)(J^-_{-1})^n\ket{j,j}\\
=& \delta^{a,-}n(k-2j+n+1)\bigg[f_{n,n-1}\bra{j,j}V^{sl}_{-1/2}(z_1)V^{sl}_{-1/2}(z_2)J^-_{-1}\ket{j,j}\\
&+(n-1)f_{n,n-2}(z_1t_1^-+z_2t_2^-)\bra{j,j}V^{sl}_{-1/2}(z_1)V^{sl}_{-1/2}(z_2)(J^-_{-1})^2\ket{j,j}\\
&+(n-1)(n-2)f_{n,n-3}z_1z_2t^-_1t^-_2\bra{j,j}V^{sl}_{-1/2}(z_1)V^{sl}_{-1/2}(z_2)(J^-_{-1})^3\ket{j,j}\bigg]\\
=&  \delta^{a,-}n(k-2j+n+1)\bigg[-f_{n,n-1}(z_1^{-1}t_1^++z_2^{-1}t_2^+)+2(n-1)f_{n,n-2}(z_1t_1^-+z_2t_2^-)z_1^{-1}z_2^{-1}t_1^+t_2^+\bigg]\\
&\times\bra{j,j}V^{sl}_{-1/2}(z_1)V^{sl}_{-1/2}(z_2)\ket{j,j},
\fe
where we have used $\bra{j,j}V^{sl}_{-1/2,m_1}(z_1)V^{sl}_{-1/2,m_2}(z_2)(J^-_{-1})^n\ket{j,j}=0$ for $n>2$. The remaining terms are handled similarly. In the end, we find
\ie\label{A2def}
X^a =& (z_{12}^{-1}(t_2^a)^T-\delta^{a,+}z_1^{-1}(t_1^++t_2^+)+\delta^{a,3}z_1^{-1}(j-n))W\\
&+\bigg\{\delta^{a,-}n(k-2j+n+1)\bigg[-f_{n,n-1}(z_1^{-1}t_1^++z_2^{-1}t_2^+)+2(n-1)f_{n,n-2}(z_1t_1^-+z_2t_2^-)z_1^{-1}z_2^{-1}t_1^+t_2^+\bigg]\\
&+\delta^{a,+}z_1^{-2}n(k-2j+n+1)\bigg[f_{n,n-1}(z_1t_1^-+z_2t_2^-)-2(n-1)f_{n,n-2}(z_1^{-1}t^+_{1}+z_2^{-1}t_2^+)z_1z_2t_1^-t_2^-\bigg]\bigg\}\\
&\times\bra{j,j}V^{sl}_{-1/2}(z_1)V^{sl}_{-1/2}(z_2)\ket{j,j}
\\
\equiv&\mathbb{X}^a(z_1,z_2)\bra{j,j}V^{sl}_{-1/2}(z_1)V^{sl}_{-1/2}(z_2)\ket{j,j},
\fe
where we again defined a matrix $\mathbb{X}^a(z_1,z_2)$ acting on the holomorphic spinor indices, and the anti-holomorphic analogue is similarly defined.

The relevant $SU(2)_{k-2}$ correlator is much simpler to compute. The correlator of interest is
\ie
(Y^{a'})_{\alpha'\beta'} \equiv & \langle V^{su}_{j',j'}(\infty)K^{a'}_{-1}V^{su}_{1/2,\alpha'}(z_1)V^{su}_{1/2,\beta'}(z_2)V^{su}_{j',j'}(0)\rangle.
\fe
One easily finds
\ie\label{Bdef}
Y^{a'}=& \bigg(z_{12}^{-1}(T_2^{a'})^T + \delta^{{a'},3}z_1^{-1}j'-\delta^{{a'},-}z_1^{-1}(T_1^++T_2^+)\bigg)\bra{j',j'}V^{su}_{1/2}(z_1)V^{su}_{1/2}(z_2)\ket{j',j'}\\
\equiv&\mathbb{Y}^{a'}(z_1,z_2)\bra{j',j'}V^{su}_{1/2}(z_1)V^{su}_{1/2}(z_2)\ket{j',j'},
\fe
where now $\ket{j',j'} = V^{su}_{j',j'}(0)\ket{0}$ and $T^a_i$ are the $SU(2)$ generator matrices acting on the current algebra primaries at $z_i$.

\section{Four point functions in the $SU(2)_{k-2}$ WZW model}\label{SU4pt} 
We review the four point function of current primaries in the $SU(2)_{k-2}$ WZW model, computed in \cite{Zamolodchikov:1986bd}. In particular, we take one of the spins to be ${1\over2}$ and solve the Knizhnik-Zamolodchikov (KZ) equation \cite{Knizhnik:1984nr}. For convenience, we define
\ie
V_j(x,\bar x;z, \bar z)=\sum_{m,\bar m=-j}^j \sqrt{{2j\choose m+j}~{2j\choose \bar m+j}}x^{j+m}{\bar x}^{j+\bar m}V_{j,m,\bar m}(z,\bar z),
\fe
where ${N\choose k}$ is the binomial coefficient. We write the four point function of interest as
\ie
G(j_i;x_i,\bar x_i;z_i,\bar z_i)=&|z_{12}^{-2h_1}z_{24}^{h_1+h_3-h_2-h_4}z_{23}^{h_1+h_4-h_2-h_3}z_{34}^{h_2-h_1-h_3-h_4}x_{13}x_{24}^{{1\over2}-j_3+j_2+j_4}x_{23}^{-{1\over2}-j_4+j_2+j_3}x_{34}^{-{1\over2}-j_2+j_3+j_4}|^2
\\
&\times V(x,\bar x;z, \bar z),
\fe
where we set $j_1={1\over2}$, $h_i={j_i(j_i+1)\over k}$, $z={z_{12}z_{34}\over z_{13}z_{24}},~x={x_{12}x_{34}\over x_{13}x_{24}}$. We fix $z_1=z, z_2=0, z_3=\infty, z_4=1$ and similarly for $x$. Due to $j_1=1/2$, $V$ has the expansion $V(x,z)=V_0(z)+x V_1(z)$, where antiholomorphic labels are omitted. The KZ equation for $j_1=1/2$ reads
\ie{}
&-kz(z-1)V_0'(z)+\left(-{3\over2}-j_2+(1+j_3)z\right)V_0(z)+{1\over2}(1-2j_2+2j_3+2j_4)zV_1(z)=0
\\
&-kz(z-1)V_1'(z)+\left(j_2-{1\over2}-j_3z\right)V_1(z)+\left({1\over2}+j_2-j_3+j_4\right)V_0(z)=0.
\fe
Writing $V_1(z)=f(z)g(z)$ with $f(z)=z^{1-2j_2\over2k}(1-z)^{2j_2-2j_3-1\over2k}$, these reduce to
\ie{}
&V_0={-1\over{1\over2}+j_2-j_3+j_4}kz(1-z)z^{1-2j_2\over2k}(1-z)^{2j_2-2j_3-1\over2k}g'(z),
\\
&g(z)=(1-z)^{1-2j_2+2j_3+2j_4\over2k}p(z),
\\
&z(1-z)p''+(c-(a+b+1)z)p'-abp=0,
\\
&a={2k-1-2j_2-2j_3+2j_4\over2k},~b={1-2j_2+2j_3+2j_4\over2k},~c={k-2j_2-1\over k}.
\fe
Therefore, $p(z)$ is the hypergeometric function ${}_2F_1(a,b,c,z)$ or $z^{1-c}{}_2F_1(1+a-c,1+b-c,2-c,z)$. Defining $F=z^{-{3\over2k}}V$, we have the following two linearly independent solutions for $F$
\ie
F^-(x,z)=&-{b\over c}z^{-j_2-1+k\over k}(1-z)^{j_4\over k}{}_2F_1(a,b+1,c+1,z)+xz^{-j_2-1\over k}(1-z)^{j_4\over k}{}_2F_1(a,b,c,z)
\\
F^+(x,z)=&-{c-1\over c-a}z^{j_2\over k}(1-z)^{j_4\over k}{}_2F_1(a-c,b-c+1,1-c,z)
\\
&+xz^{j_2\over k}(1-z)^{j_4\over k}{}_2F_1(1+a-c,1+b-c,2-c,z).
\fe
From the weights, it is clear that $F^-$ is the $\left(j_2-{1\over2}\right)$ block while $F^+$ is the $\left(j_2+{1\over2}\right)$ block. From single-valuedness of the four point function, we can only take diagonal combinations
\ie
G(j_i;x,\bar x;z, \bar z)=\sum_{\sigma=\pm}s_\sigma F^\sigma(x,z)F^\sigma(\bar x,\bar z).
\fe
We fix $s_\sigma$ by crossing and identity channel normalization. We denote the above blocks $F^\sigma$ as $F^\sigma_{24}$, indicating the labels $j_2$ and $j_4$. Under crossing, we also need to exchange these labels. Thus, crossing takes $z\rightarrow1-z,x\rightarrow1-x$ and at the same time $j_2\leftrightarrow j_4$, which amounts to $a\rightarrow1-b,b\rightarrow1-a,c\rightarrow1-a-b+c$.
\ie
F^\sigma_{24}(x,z)=\sum_\rho e_{\sigma\rho}F^\rho_{42}(1-x,1-z).
\fe
To get the $e_{\sigma\rho}$ coefficients, we use the following identity
\ie
{}_2F_1(a,b,c,1-z)=&{\Gamma(a+b-c)\Gamma(c)\over\Gamma(a)\Gamma(b)}z^{c-a-b}{}_2F_1(c-a,c-b,c-a-b+1,z)
\\
&+{\Gamma(c-a-b)\Gamma(c)\over\Gamma(c-a)\Gamma(c-b)}{}_2F_1(a,b,a+b-c+1,z),
\fe
and write
\ie
F^-_{24}(x,z)&=z^{-j_2-1\over k}(1-z)^{j_4\over k}\left(e_{--}{\Gamma(1-c)\Gamma(1-a-b+c)\over\Gamma(1-a)\Gamma(1-b)}+e_{-+}{\Gamma(1-c)\Gamma(a+b-c+1)\over\Gamma(a-c+1)\Gamma(b-c+1)} \right)
\\
&\times\left({}_2F_1(a,b,c,z)-{}_2F_1(a-1,b,c,z) \right)
\\
&+z^{j_2\over k}(1-z)^{-j_4-1\over k}\left(e_{--}{\Gamma(c-1)\Gamma(1-a-b+c)\over\Gamma(c-a)\Gamma(c-b)}+e_{-+}{\Gamma(a+b-c+1)\Gamma(c-1)\over\Gamma(a)\Gamma(b)} \right)
\\
&\times{}_2F_1(1-a,1-b,2-c,z)
\\
&-z^{j_2\over k}(1-z)^{-j_4-1+k\over k}\left(e_{--}{(1-a)\Gamma(c-1)\Gamma(1-a-b+c)\over\Gamma(c-a+1)\Gamma(c-b)}+e_{-+}{\Gamma(a+b-c+1)\Gamma(c-1)\over(a-c)\Gamma(a-1)\Gamma(b)} \right)
\\
&\times{}_2F_1(2-a,1-b,2-c,z)
\\
&-xz^{j_2\over k}(1-z)^{-j_4-1\over k}\left(e_{--}{\Gamma(c-1)\Gamma(1-a-b+c)\over\Gamma(c-a)\Gamma(c-b)}+e_{-+}{\Gamma(c-1)\Gamma(a+b-c+1)\over\Gamma(a)\Gamma(b)} \right)
\\
&\times{}_2F_1(1-a,1-b,2-c,z)
\\
&-xz^{-j_2-1\over k}(1-z)^{j_4\over k}\left(e_{--}{\Gamma(1-c)\Gamma(1-a-b+c)\over\Gamma(1-a)\Gamma(1-b)}+e_{-+}{\Gamma(a+b-c+1)\Gamma(1-c)\over\Gamma(a-c+1)\Gamma(b-c+1)} \right)
\\
&\times{}_2F_1(a,b,c,z),
\fe
where, for $F^+_{24}(x,z)$, we replace $e_{--}\rightarrow e_{+-}$ and $e_{-+}\rightarrow e_{++}$ in the above expression. Using ${}_2F_1(a,b,c,z)-{}_2F_1(a-1,b,c,z)={b\over c}z{}_2F_1(a,b+1,c+1,z)$ and ${}_2F_1(1+a-c,1+b-c,2-c,z)-{a-1\over a-c}{}_2F_1(a-c,1+b-c,2-c,z)={1-c\over a-c}{}_2F_1(a-c,1+b-c,1-c,z)$ to compare with the definition of $F^{\pm}_{24}(x,z)$, we get
\ie
{}&{e_{--}\over e_{-+}}=-{\Gamma(a+b-c+1)\Gamma(c-a)\Gamma(c-b)\over\Gamma(1-a-b+c)\Gamma(a)\Gamma(b)}
\\
&e_{-+}\Gamma(1-c)\Gamma(a+b-c+1)\left({\Gamma(c-a)\Gamma(c-b)\over\Gamma(a)\Gamma(b)\Gamma(1-a)\Gamma(1-b)}-{1\over\Gamma(a-c+1)\Gamma(b-c+1)} \right)=1,
\\
&{e_{+-}\over e_{++}}=-{\Gamma(a+b-c+1)\Gamma(1-a)\Gamma(1-b)\over\Gamma(1-a-b+c)\Gamma(a-c+1)\Gamma(b-c+1)},
\\
&e_{++}\Gamma(c-1)\Gamma(a+b-c+1)\left({\Gamma(1-a)\Gamma(1-b)\over\Gamma(a-c+1)\Gamma(b-c+1)\Gamma(c-a)\Gamma(c-b)}-{1\over\Gamma(a)\Gamma(b)}\right)=1.
\fe
Crossing symmetry of the four point function
\ie
G(j_i;x,\bar x;z, \bar z)&=\sum_{\sigma=\pm}s_\sigma^{24} F^\sigma_{24}(x,z)F^\sigma_{24}(\bar x,\bar z)=\sum_{\sigma=\pm}s_\sigma^{24}\sum_{\rho,\bar\rho}e_{\sigma\rho}e_{\sigma\bar\rho} F^\rho_{42}(1-x,1-z)F^{\bar\rho}_{42}(1-\bar x,1-\bar z)
\\
&=\sum_{\sigma=\pm}s_\sigma^{42} F^\sigma_{42}(1-x,1-z)F^\sigma_{42}(1-\bar x,1-\bar z),
\fe
gives the following relations
\ie
s_+^{24}=-{e_{--}e_{-+}\over e_{+-}e_{++}}s^{24}_-,~~~s_-^{42}=\left(e_{--}^2-{e_{--}e_{-+}e_{+-}\over e_{++}}\right)s_-^{24}.
\fe
The normalization for the case $j_4={1\over2}$ is given by the limit $z\sim1$ where the identity block dominates
\ie
F^-_{42}(1-x,1-z)=(1-z)^{-{3\over2k}}-x(1-z)^{-{3\over2k}}+...
\fe
Therefore, we deduce
\ie
s_-^{42}=1~~~\Rightarrow ~~~s_-^{24}={e_{++}\over e_{--}}~{1\over e_{--}e_{++}-e_{-+}e_{+-}},~~s^{24}_+=-{e_{-+}\over e_{+-}}~{1\over e_{--}e_{++}-e_{-+}e_{+-}}.
\fe
With these, the final answer is given by
\ie
G(j_i;x,\bar x;z, \bar z)=s_-^{24}F^-_{24}(x,z)F^-_{24}(\bar x,\bar z)+s_+^{24}F^+_{24}(x,z)F^+_{24}(\bar x,\bar z).
\fe
For our case of interest, we set $j_2=j_3=j'$ and $j_4={1\over2}$. Then,
\ie
a=1-{2j'\over k},~~b={1\over k},~~c=1-{2j'+1\over k}.
\fe
For $j'={1\over2}$, we can read off the OPE coefficient for $SU(2)$ primaries with $j'$ values ${1\over2},{1\over2},1$
\ie
\left(C^{SU(2)}_{1/2,1/2,1}\right)^2=4s_+^{24}\bigg |_{a=1-1/k,b=1/k,c=1-2/k},
\fe
which agrees with the result in \cite{Zamolodchikov:1986bd}. Taking $z_2=0$ and $z_3=\infty$,
\ie
G(j_i;x_i,\bar x_i;z_i, \bar z_i)=|z_4^{-{3\over2k}}x_{13}x_{24}x_{23}^{2j-1}|^2\left[s_-^{24}F^-_{24}(x,z_1/z_4)F^-_{24}(\bar x,{\bar z}_1/{\bar z}_4)+s_+^{24}F^+_{24}(x,z_1/z_4)F^+_{24}(\bar x,{\bar z}_1/{\bar z}_4) \right].
\fe
Finally, the correlator of interest is
\ie\label{sufourpt}
G^{su,j'}_{m_1,\bar{m}_1;m_2,\bar{m}_2}(z_i,\bar{z_i})&=\langle V_{j',-j',-j'}(\infty)V_{j',j',j'}(0)V_{1/2,m_1,\bar{m}_1}(z_1,\bar{z}_1)V_{1/2,m_2,\bar{m}_2}(z_2,\bar{z}_2)\rangle\\
&=|z_2|^{-{3\over k}}g^{su,j'}_{m_1,\bar{m}_1;m_2,\bar{m}_2}(z_1/z_2,\bar{z_1}/\bar{z_2}),
\fe
with
\ie\label{su4pt}
&g^{su,j'}_{++;--}(z,\bar z)=s_-^{24}F_1(z)F_1(\bar z)+s_+^{24}F_3(z)F_3(\bar z)
\\
&g^{su,j'}_{+-;-+}(z,\bar z)=s_-^{24}F_1(z)F_2(\bar z)+s_+^{24}F_3(z)F_4(\bar z)
\\
&g^{su,j'}_{-+;+-}(z,\bar z)=s_-^{24}F_2(z)F_1(\bar z)+s_+^{24}F_4(z)F_3(\bar z)
\\
&g^{su,j'}_{--;++}(z,\bar z)=s_-^{24}F_2(z)F_2(\bar z)+s_+^{24}F_4(z)F_4(\bar z).
\fe
The functions $F_i$ above, $i=1,2,3,4$, are given by
\ie{}
&F_1(z)={1\over 2j'+1-k}z^{1-{j'+1\over k}}(1-z)^{1\over 2k}{}_2F_1\left(1-{2j'\over k},1+{1\over k},2-{2j'+1\over k},z\right)
\\
&F_2(z)=z^{-{j'+1\over k}}(1-z)^{1\over 2k}{}_2F_1\left(1-{2j'\over k},{1\over k},1-{2j'+1\over k},z\right)
\\
&F_3(z)=-(2j'+1)z^{j'\over k}(1-z)^{1\over 2k}{}_2F_1\left({1\over k},{2j'+2\over k},{2j'+1\over k},z\right)
\\
&F_4(z)=z^{j'\over k}(1-z)^{1\over 2k}{}_2F_1\left(1+{1\over k},{2j'+2\over k},1+{2j'+1\over k},z\right).
\fe

\section{Four point functions in the $SL(2,\mathbb{R})_{k+2}$ WZW model}\label{SL4pt} 
The computation for $SL(2,\mathbb{R})$ is analogous since we consider four point functions of two doublets and two discrete representations
\ie
G(y,\bar y; z,\bar z)&=\sum_{l,\bar l=0,1}y^l{\bar y}^{\bar l}G_{l,\bar l}(z,\bar z)
\\
&=\sum_{l,\bar l=0,1}y^l{\bar y}^{\bar l}\langle V_{-{1\over2},{1\over2}-l,{1\over2}-\bar l}(z,\bar z)V_{j_2,j_2,j_2}(0,0)V_{j_3,-j_3,-j_3}(\infty,\infty)V_{j_4,j_4+l,j_4+\bar l}(1,1)  \rangle,
\fe
where $j_4$ will later be taken to be $-{1\over2}$ corresponding to a doublet. For the KZ equation, focusing on just the holomorphic part, we have
\ie{}
&z(z-1)k\partial_z G_0-(z-1)(2j_3-3j_2-2j_4)G_0+(z-1)i\sqrt{2j_4}G_1+zj_4G_0-zi\sqrt{2j_4}G_1=0
\\
&z(z-1)k\partial_z G_1-(z-1)j_2G_1-z(1+j_4)G_1-zi\sqrt{2j_4}G_0=0.
\fe
Taking $G_1\sim z^A(1-z)^Bp(z)$ with $A={j_2\over k},B={j_4\over k}$, $p(z)$ satisfies the hypergeometric differential equation
\ie
&p(z)={}_2F_1(a,b,c,z)~~\text{or}~~z^{1-c}{}_2F_1(1+a-c,1+b-c,2-c,z),
\\
&a=-{1\over k}, ~b={2(2j_2-j_3+2j_4)\over k}, ~c={2(2j_2-j_3+j_4)\over k}.
\fe
Accordingly, the solution for $G_0$ is given by
\ie
G_0\sim{ik\over\sqrt{2j_4}}z^A(1-z)^{1-a+B}\partial_z((1-z)^ap(z)).
\fe
Then, the two linearly independent solutions for $G(y,z)$ are given by
\ie
F_{24}^-(y,z)=&{ia\sqrt{2kB}\over c}z^A(1-z)^B{}_2F_1(a+1,2B+c,c+1,z)+yz^A(1-z)^B{}_2F_1(a,b,c,z),
\\
F_{24}^+(y,z)=&{ik(1-c)\over\sqrt{2kB}}z^{A-c}(1-z)^B{}_2F_1(2B,1+a-c,1-c,z)
\\&+yz^{A-c+1}(1-z)^B{}_2F_1(1+a-c,1+2B,2-c,z).
\fe
Here, $F^-$ is the $(j_2-{1\over2})$ block, while $F^+$ generically does not have the form of a specific block. For $j_3=j_2$ and $j_4=-{1\over2}$, $F^+$ corresponds to the $(j_2+{1\over2})$ block. The final answer will be of the form
\ie
G(y,\bar y;z, \bar z)=\sum_{\sigma=\pm}h_\sigma F_{24}^\sigma(y,z)F_{24}^\sigma(\bar y,\bar z),
\fe
where we take diagonal combination for single-valuedness of the correlator. To fix the coefficients $h_\sigma$, we study the transformation of the blocks under crossing. It suffices to examine the $y=0$ component. As before, we not only take $z\rightarrow 1-z$, but also take $j_2\leftrightarrow j_4$. This amounts to $A\leftrightarrow B$ and $c\rightarrow c+2B-2A$. Writing
\ie
F^\sigma_{24}(y=0,z)=\sum_\rho d_{\sigma\rho}F^\rho_{42}(y=0,1-z),
\fe
we get $d_{\sigma\rho}$ from the following
\ie
&F^-_{24}(y=0,z)=d_{--}F^-_{42}(y=0,1-z)+d_{-+}F^+_{42}(y=0,1-z)
\\
&=d_{--}{ia\sqrt{2kA}\over c+2B-2A}z^A(1-z)^B\bigg({\Gamma(-a-2A)\Gamma(1+c-2A+2B)\over\Gamma(1-2A)\Gamma(c-a-2A+2B)}{}_2F_1(1+a,c+2B,1+a+2A,z)
\\
&+z^{-a-2A}{\Gamma(a+2A)\Gamma(1+c-2A+2B)\over\Gamma(1+a)\Gamma(c+2B)}{}_2F_1(1-2A,-a+c-2A+2B,1-a-2A,z)\bigg)
\\
&+d_{-+}{ik(1-c+2A-2B)\over\sqrt{2kA}}z^A(1-z)^{2A-B-c}
\\
&\times\bigg({\Gamma(a+2A)\Gamma(1-c+2A-2B)\over\Gamma(2A)\Gamma(1+a-c+2A-2B)}z^{-a-2A}{}_2F_1(-a,1-c-2B,1-a-2A,z)
\\
&+{\Gamma(-a-2A)\Gamma(1-c+2A-2B)\over\Gamma(-a)\Gamma(1-c-2B)}{}_2F_1(2A,1+a-c+2A-2B,1+a+2A,z)\bigg),
\fe
and similarly for $F^+_{24}(y=0,z)$. This relation cannot be satisfied for generic values of parameters, meaning that there is no such crossing kernel, but it can be satisfied for our specific case where $j_2=j_3=j$ and $j_4=-{1\over2}$. For this case, the parameters are
\ie
a=-{1\over k},~c={2j-1\over k},~A={j\over k},~B=-{1\over2k}.
\fe
Then, the coefficients $d_{\sigma\rho}$ are the solutions to the following equations
\ie
{}&d_{--}i\sqrt{j\over2}{\Gamma\left(1-{2\over k}\right)\Gamma\left({1-2j\over k}\right)\over\Gamma\left(1-{2j\over k}\right)\Gamma\left(-{1\over k}\right)}+d_{-+}{i(k+2)\over\sqrt{2j}}{\Gamma\left({1-2j\over k}\right)\Gamma\left(1+{2\over k}\right)\over\Gamma\left(1+{2-2j\over k}\right)\Gamma\left({1\over k}\right)}={1\over 2j-1},
\\
&d_{--}j{\Gamma\left(1-{2\over k}\right)\over\Gamma\left({2j-2\over k}\right)\Gamma\left(1-{1\over k}\right)}+d_{-+}(k+2){\Gamma\left(1+{2\over k}\right)\over\Gamma\left(1+{1\over k}\right)\Gamma\left({2j\over k}\right)}=0,
\\
&d_{+-}i\sqrt{j\over2}{\Gamma\left({2j-1\over k}\right)\Gamma\left(1-{2\over k}\right)\over\Gamma\left(1-{1\over k}\right)\Gamma\left({2j-2\over k}\right)}+d_{++}{i(k+2)\over\sqrt{2j}}{\Gamma\left({2j-1\over k}\right)\Gamma\left(1+{2\over k}\right)\over\Gamma\left({2j\over k}\right)\Gamma\left(1+{1\over k}\right)}=k-2j+1,
\\
&d_{+-}j{\Gamma\left(1-{2\over k}\right)\over\Gamma\left(1-{2j\over k}\right)\Gamma\left(-{1\over k}\right)}+d_{++}(k+2){\Gamma\left(1+{2\over k}\right)\over\Gamma\left({1\over k}\right)\Gamma\left(1-{2j-2\over k}\right)}=0.
\fe
With these, $h_\sigma$ coefficients are given by
\ie
h_-^{24}=-{d_{+-}d_{++}\over d_{--}d_{-+}}h_+^{24},~~~h_+^{42}=\left(d_{++}^2-{d_{+-}d_{++}d_{-+}\over d_{--}} \right)h^{24}_+,
\fe
and we can fix $h_+^{42}$ by normalization of two point function. The identity channel is given by the following block near $z\sim1$
\ie
F^+_{42}(y=0,1-z)={i(k+2)\over\sqrt{2j}}(1-z)^{3\over2k}+...
\fe
Therefore, we get
\ie
h_+^{42}=-{2j\over(k+2)^2},
\fe
which then determines $h_\sigma^{24}$. The OPE coefficient for three current algebra primaries carrying $j$ values $-{1\over2},-{1\over2},-1$ is given by the $(\sigma=-)$ block, leading to (\ref{OPEcoeffs}). In the end, the correlator of interest is
\ie\label{slfourpt}{}
G^{sl,j}_{m_1,\bar{m}_1;m_2,\bar{m}_2}(z_i,\bar{z_i})=&\langle V_{j,-j,-j}(\infty)V_{j,j,j}(0)V_{-1/2,m_1,\bar{m}_1}(z_1,\bar{z_1})V_{-1/2,m_2,\bar{m}_2}(z_2,\bar{z_2})\rangle
\\
=&|z_2|^{3\over k}g^{sl,j}_{m_1,\bar{m}_1;m_2,\bar{m}_2}(z_1/z_2,\bar{z_1}/\bar{z_2}),
\fe
with
\ie\label{sl4pt}
&g^{sl,j}_{++;--}(z,\bar z)=h^{24}_-L_1(z)L_1(\bar z)+h^{24}_+L_3(z)L_3(\bar z)
\\
&g^{sl,j}_{+-;-+}(z,\bar z)=h^{24}_-L_1(z)L_2(\bar z)+h^{24}_+L_3(z)L_4(\bar z)
\\
&g^{sl,j}_{-+;+-}(z,\bar z)=h^{24}_-L_2(z)L_1(\bar z)+h^{24}_+L_4(z)L_3(\bar z)
\\
&g^{sl,j}_{--;++}(z,\bar z)=h^{24}_-L_2(z)L_2(\bar z)+h^{24}_+L_4(z)L_4(\bar z).
\fe
The functions $L_i$ above, $i=1,2,3,4$, are given by
\ie
&L_1(z)={1\over2j-1}z^{j\over k}(1-z)^{-{1\over2k}}{}_2F_1\left(1-{1\over k},{2j-2\over k},1+{2j-1\over k},z\right)
\\
&L_2(z)=z^{j\over k}(1-z)^{-{1\over2k}}{}_2F_1\left(-{1\over k},{2j-2\over k},{2j-1\over k},z\right)
\\
&L_3(z)=(k+1-2j)z^{1-j\over k}(1-z)^{-{1\over2k}}{}_2F_1\left(-{1\over k},1-{2j\over k},1-{2j-1\over k},z\right)
\\
&L_4(z)=z^{1+{1-j\over k}}(1-z)^{-{1\over2k}}{}_2F_1\left(1-{1\over k},1-{2j\over k},2-{2j-1\over k},z\right).
\fe

\bibliographystyle{JHEP}
\bibliography{RRSFTv4}

\end{document}